\documentclass[aps,prd,twocolumn,superscriptaddress,10pt,nofootinbib, floatfix]{revtex4-2} 
\usepackage{tabularx}
\usepackage{graphicx}
\usepackage{amsmath}
\usepackage{amssymb}
\usepackage{comment}
\usepackage{soul} 
\usepackage[dvipsnames]{xcolor}
\usepackage{xspace}
\usepackage{multirow}
\usepackage{natbib}
\usepackage{hyperref}
\hypersetup{hidelinks}
\usepackage{url}
\usepackage{booktabs}
\usepackage{multirow}
\usepackage[normalem]{ulem}
\usepackage{orcidlink}
\usepackage[caption=false]{subfig} 
\usepackage{graphicx}
\usepackage[normalem]{ulem}
\usepackage{cancel}
\usepackage{xcolor}

\newcommand\apjl{The Astrophysical Journal Letters}

\renewcommand{\selectlanguage}[1]{}

\begin{document}

\title{Red noise and evolving signals: a complete frequentist approach to supermassive black hole binary searches with pulsar timing arrays}

\author{Xuan Tao\orcidlink{0009-0000-6875-7411}}
\affiliation{National Gravitation Laboratory, MOE Key Laboratory of Fundamental Physical Quantities Measurements, Department of Astronomy and School of Physics, Huazhong University of Science and Technology, Wuhan 430074, China}
\affiliation{Max Planck Institute for Gravitational Physics (Albert Einstein Institute), 30167 Hannover, Germany}
\affiliation{Leibniz Universität Hannover, 30167 Hannover, Germany}

\author{Boris Goncharov\orcidlink{0000-0003-3189-5807}}
\affiliation{Max Planck Institute for Gravitational Physics (Albert Einstein Institute), 30167 Hannover, Germany}
\affiliation{Leibniz Universität Hannover, 30167 Hannover, Germany}

\author{Yiqian Qian\orcidlink{0000-0002-8039-5617}}
\affiliation{National Gravitation Laboratory, MOE Key Laboratory of Fundamental Physical Quantities Measurements, Department of Astronomy and School of Physics, Huazhong University of Science and Technology, Wuhan 430074, China}

\author{Yan Wang\orcidlink{0000-0001-8990-5700}}
\email{ywang12@hust.edu.cn}
\affiliation{National Gravitation Laboratory, MOE Key Laboratory of Fundamental Physical Quantities Measurements, Department of Astronomy and School of Physics, Huazhong University of Science and Technology, Wuhan 430074, China}

\author{Soumya D.~Mohanty\orcidlink{0000-0002-4651-6438}}
\email{soumya.mohanty@utrgv.edu}
\affiliation{Department of Physics and Astronomy, University of Texas Rio Grande Valley, Brownsville, Texas 78520, USA}
\affiliation{Department of Physics, IIT Hyderabad, Kandai, Telangana-502284, India}

\begin{abstract}
Searches for gravitational waves (GWs) from isolated supermassive black hole binaries (SMBHBs) in pulsar timing array (PTA) data require simultaneous estimation of signal and noise parameters, so the dimensionality of the fit scales with the number of observed pulsars. This computational difficulty is exacerbated when source evolution from GW emission is included, since retaining both Earth and pulsar terms introduces the unknown pulsar distances. 
Existing frequentist methods such as the \(\mathcal F\)-statistic are restricted to non-evolving sources. In addition, they often rely on a noise covariance estimated from the same data and then held fixed during the signal search, which can bias parameter estimates.
We present a Generalized Likelihood Ratio Test and the associated $\mathcal{T}$-statistic that overcomes the aforementioned limitations. This formulation extends earlier work in which the dimensionality of the fitting problem was drastically reduced by semi-analytical maximization of the likelihood over the pulsar phase parameters, followed by efficient global optimization over the remaining parameters using Particle Swarm Optimization. Our simulations demonstrate that for an evolving SMBHB signal with chirp mass $\mathcal{M}=10^{9.2}\,M_\odot$ and signal-to-noise ratio $20$, this detection statistic achieves a $100\%$ detection probability at a false-alarm probability of $0.06$ in a 30-pulsar timing array, which is characterized by a $100~\mathrm{ns}$ root-mean-square white noise residual and pulsar-specific red noise. 
For the 30-pulsar timing array at signal-to-noise ratio $10$ and false-alarm probability $0.06$, $\mathcal{T}$ detects $99/100$ realizations, outperforming the $\mathcal{F}_p$ statistic evaluated with the $H_0$-fitted covariance, which detects $71/100$ realizations.
\end{abstract}

\maketitle

\section{Introduction}
The detection of nanohertz-frequency continuous gravitational waves (CGWs) emitted by individual supermassive black hole binaries (SMBHBs) remains a primary scientific objective for Pulsar Timing Array (PTA) collaborations \cite{Sazhin_1978, Detweiler_1979, Foster_1990, Hobbs_2010, BurkeSpolaor_2019}. Following milestone announcements by the North American Nanohertz Observatory for Gravitational Waves (NANOGrav) \cite{Agazie_2023_nanograv}, the European Pulsar Timing Array (EPTA) \cite{Antoniadis_2023_epta}, the Parkes Pulsar Timing Array (PPTA) \cite{Reardon_2023_ppta}, the Chinese Pulsar Timing Array (CPTA) \cite{Xu_2023_cpta}, and the Indian Pulsar Timing Array (InPTA) \cite{Tarafdar_2022_inpta} regarding evidence of a stochastic gravitational-wave background (GWB) exhibiting the signature Hellings-Downs spatial correlation \cite{Hellings_1983}, the focus of the community is rapidly expanding toward resolving individual, deterministic CGW sources emerging above this background \cite{Rajagopal_1995, Jaffe_2003, Sesana_2008, Kelley_2017}. Detecting a single SMBHB would allow for the detailed study of binary orbital dynamics, environmental coupling mechanisms, and the extraction of system parameters \cite{Mingarelli_2012}. 
Furthermore, staggered sampling, an inherent feature of PTA data, may allow PTAs to extend their frequency reach into the microhertz regime~\cite{2021ApJ...907L..43W}, thereby enabling the detection of ringdown signals from SMBHB mergers~\cite{2025PhRvD.111h3030T}.

Searching for individual CGW sources in time-domain PTA data, however, poses a formidable data analysis challenge due to the composite nature of the signal. The induced timing residuals from a CGW consist of two components: the Earth term, representing the metric perturbation as it passes the Solar System Barycenter (SSB), and the pulsar term, representing the perturbation at the pulsar \cite{Corbin_2010, Lee_2011}.  While the Earth term is coherent across all pulsars in the array, the pulsar term is delayed by the light-travel time from the pulsar to Earth. Because pulsar distances are typically known with large uncertainties \cite{Deller_2008, Verbiest_2012}, the exact phase of the gravitational wave at each pulsar acts as a random variable. For an array of $N_p$ pulsars, fully modeling the signal requires exploring a high-dimensional parameter space containing the global binary parameters plus $2N_p$ additional pulsar-specific parameters, rendering traditional grid-based or naive stochastic searches computationally intractable.

Historically, this dimensionality has often been reduced by analytically maximized detection statistics. Techniques first developed for ground-based interferometers \cite{Jaranowski_1998} were later adapted to PTAs, most notably as the $\mathcal{F}_e$-statistic and the $\mathcal{F}_p$-statistic \cite{Babak_2012, Ellis_2012, Zhu_2014}. These statistics are attractive because they are easy to implement and, under fixed-noise assumptions, lead to simple reference distributions for threshold setting \cite{Arzoumanian_2014}. However, this simplicity comes from approximations. The coherent $\mathcal{F}_e$-statistic omits the pulsar term to preserve phase coherence, while the $\mathcal{F}_p$-statistic includes the pulsar term only through an incoherent maximization over pulsar-dependent amplitude parameters. Standard implementations also assume a monochromatic source, neglecting radiation-driven frequency evolution during the observing span \cite{Corbin_2010}.

Beyond these waveform-level approximations, existing frequentist CW statistics also face a noise-model limitation. The standard PTA $\mathcal{F}_e$ and $\mathcal{F}_p$ statistics are covariance-weighted likelihood-ratio statistics for irregularly sampled timing residuals, with timing-model fitting and colored noise included in the inner product \cite{Babak_2012,Ellis_2012,Zhu_2014,Arzoumanian_2014}. Their analytic null distributions are therefore obtained only after the noise covariance has been specified. In real PTA data, however, this covariance includes pulsar-specific red noise, chromatic propagation effects, and other stochastic components that must be inferred from the data \cite{van_haasteren_new_2014,Ellis_2019_enterprise}.

This creates a practical limitation for frequentist CW searches. Current analyses can use Bayesian sampling to infer the relevant noise and signal parameters, and then either carry out a fully Bayesian CW search or evaluate frequentist summary statistics using posterior-informed noise covariance information \cite{Vigeland_2018,aggarwal_nanograv_2019,Arzoumanian_2023_cw,Becsy_2022,sardesai_optimal_2026}. These strategies are powerful, but they also move away from the original clean frequentist setting: the statistic is no longer simply a fixed-covariance $\mathcal{F}$-statistic with an analytic reference distribution, and it is not a profile-likelihood ratio in which the nuisance covariance is optimized separately under the signal and noise-only hypotheses. When a low-frequency CGW signal is covariant with intrinsic red noise, this distinction can reduce sensitivity because part of the deterministic signal power may be absorbed into the effective noise covariance.

Motivated by these limitations and using the Generalized Likelihood Ratio Test (GLRT) formalism~\cite{kay1998fundamentals}, we present a time-domain search that jointly fits the global CGW parameters and the red-noise parameters of individual pulsars. The full PTA likelihood is maximized under the signal hypothesis $H_1$, including both the CGW and red-noise parameters, and independently under the noise-only hypothesis $H_0$, including only the red-noise parameters. The two hypotheses are therefore compared using their separately profiled covariance matrices rather than a single externally supplied or posterior-averaged covariance matrix. The method incorporates both Earth and pulsar terms and accounts for GW radiation-driven orbital frequency evolution. To handle the high dimensionality introduced by the pulsar terms, we decouple the pulsar distance from the pulsar phase and analytically maximize over the latter. The remaining parameters are optimized with a Particle Swarm Optimization (PSO) algorithm~\cite{Kennedy_1995,mohanty2018swarm}. 

The performance of the detection statistics is evaluated rigorously using simulated noise realizations. We find that, for the 30-pulsar timing array at signal-to-noise ratio $10$ and a target nominal false-alarm probability of $0.06$, $\mathcal{T}$ detects $99/100$ realizations, compared with $71/100$ for $\mathcal{F}_p$ evaluated using the $H_0$-fitted covariance and the analytic fixed-covariance threshold.

The remainder of this paper is organized as follows. In Section \ref{sec:Data model1}, we describe the mathematical formulation of the CGW data model and the decoupling of the pulsar phase. In Section \ref{sec:Methods}, we detail the PTA likelihood function, the analytical quartic maximization of the pulsar term, and the implementation of the PSO algorithm. Section \ref{sec:SimResults} presents the results of our simulated searches, focusing on the behavior of the empirical false alarm tail in two representative PTA configurations and the accuracy of our joint parameter estimation. We also compare the new method with the $\mathcal{F}_p$ statistic under different covariance choices, as a diagnostic of how noise-covariance estimation affects weak-signal performance. Finally, we conclude with a summary and discussion in Section \ref{Sec:conc}.

\section{Data model}
\label{sec:Data model1}

In this section, we review the waveforms of CGW emitted by SMBHBs and their induced delays(-advances) in pulse arrival times, explicitly including both the Earth term and the pulsar term. 
We consider circular orbits and the corresponding radiation-reaction frequency evolution. Effects of black hole spin and non-GW radiation on the orbital evolution are not considered here. 
Finally, we restructure the timing delay signal into a mathematically tractable form that isolates the highly uncertain pulsar phase as an independent extrinsic parameter, which lays the foundation for the analytical maximization described in Section~\ref{sec:Methods}. Throughout this section, we employ natural units with $G=c=1$.

In general relativity, a propagating gravitational wave is described as the superposition of two orthogonal polarization modes:
\begin{equation}
h_{a b}(t, \hat{\Omega}) = e_{a b}^{+}(\hat{\Omega}) h_{+}(t, \hat{\Omega}) + e_{a b}^{\times}(\hat{\Omega}) h_{\times}(t, \hat{\Omega})\;.
\end{equation}
The polarization tensors can be expressed as
\begin{equation}
\begin{aligned}
e_{a b}^{+}(\hat{\Omega}) &= \hat{m}_a \hat{m}_b - \hat{n}_a \hat{n}_b\;, \\
e_{a b}^{\times}(\hat{\Omega}) &= \hat{m}_a \hat{n}_b + \hat{n}_a \hat{m}_b\;,
\end{aligned}
\end{equation}
where $\hat{m}$ and $\hat{n}$ are unit vectors orthogonal to each other and to the wave propagation direction $\hat{\Omega}$. For a source located at polar angle $\theta$ and azimuthal angle $\varphi$, 
\begin{equation}
\begin{aligned}
\hat{\Omega} &= -(\sin \theta \cos \varphi) \hat{x} - (\sin \theta \sin \varphi) \hat{y} - (\cos \theta) \hat{z}\;, \\
\hat{m} &= -(\sin \varphi) \hat{x} + (\cos \varphi) \hat{y}\;, \\
\hat{n} &= -(\cos \theta \cos \varphi) \hat{x} - (\cos \theta \sin \varphi) \hat{y} + (\sin \theta) \hat{z}\;,
\end{aligned}
\end{equation}
where $\{\hat{x}, \hat{y}, \hat{z}\}$ denote the standard orthonormal basis vectors of the Solar System Barycenter (SSB) coordinate system.
The observable effect of a passing GW on a pulsar's timing residuals is the integral of the metric perturbation along the photon trajectory, which can be written as
\begin{equation}
s(t, \hat{\Omega}) = \sum_{A=+, \times} \Delta s_A(t) F^A(\hat{\Omega})\;,
\end{equation}
where
\begin{equation}
\Delta s_A(t) = s_A(t_p) - s_A(t)\;.
\end{equation}
Here, $t$ is the time the GW passes through the SSB, corresponding to the Earth term, and $t_p$ is the time it passed through the pulsar, corresponding to the pulsar term. Based on geometric considerations, $t$ and $t_p$ are related by the light-travel time:
\begin{equation}
\delta t \equiv t - t_p = L (1 + \hat{\Omega} \cdot \hat{p})\;,
\end{equation}
where $\hat{p}$ is a unit vector pointing from the SSB to the pulsar, and $L$ is the distance between the SSB and the pulsar. Furthermore, $F^A$ are the antenna pattern functions describing the geometric sensitivity of the Earth-pulsar baseline to the source, given by \cite{Ellis_2012}:
\begin{equation}
\begin{aligned}
F^{+}(\hat{\Omega}) &= \frac{1}{2} \frac{(\hat{m} \cdot \hat{p})^2 - (\hat{n} \cdot \hat{p})^2}{1 + \hat{\Omega} \cdot \hat{p}}\;, \\
F^{\times}(\hat{\Omega}) &= \frac{(\hat{m} \cdot \hat{p})(\hat{n} \cdot \hat{p})}{1 + \hat{\Omega} \cdot \hat{p}}\;.
\end{aligned}
\end{equation}
Here, the vectors $\{\hat{m}, \hat{n}\}$ and $\hat{p}$ are the polarization and pulsar direction unit vectors defined previously.

For an inspiraling circular SMBHB, at the zeroth post-Newtonian (0-PN) order, the induced residual functions $s_A$ are given by \cite{aggarwal_nanograv_2019}:
\begin{align}
s_{+}(t) &= \frac{\mathcal{M}^{5/3}}{d_L \omega(t)^{1/3}} \big[ -\sin 2\Phi(t) (1 + \cos^2 \iota) \cos 2\psi \notag \\
&\quad - 2 \cos 2\Phi(t) \cos \iota \sin 2\psi \big]\;, \\
s_{\times}(t) &= \frac{\mathcal{M}^{5/3}}{d_L \omega(t)^{1/3}} \big[ -\sin 2\Phi(t) (1 + \cos^2 \iota) \sin 2\psi \notag \\
&\quad + 2 \cos 2\Phi(t) \cos \iota \cos 2\psi \big]\;,
\end{align}
where $\psi$ is the GW polarization angle, $\iota$ is the orbital inclination angle of the SMBHB, $d_L$ is the luminosity distance to the source, and $\mathcal{M} \equiv \frac{(m_1 m_2)^{3/5}}{(m_1 + m_2)^{1/5}}$ is the chirp mass. 

Due to the continuous emission of gravitational radiation, the binary loses orbital energy, causing its frequency to increase. The orbital angular frequency $\omega(t)$ and the phase $\Phi(t)$ of the SMBHB evolve as:
\begin{equation}
\begin{aligned}
\omega(t) &= \omega_0 \left( 1 - \frac{256}{5} \mathcal{M}^{5/3} \omega_0^{8/3} t \right)^{-3/8}\;, \\
\Phi(t) &= \Phi_0 + \frac{1}{32 \mathcal{M}^{5/3}} \left( \omega_0^{-5/3} - \omega(t)^{-5/3} \right)\;.
\end{aligned}
\end{equation}
It is important to emphasize that, unlike standard implementations of the $\mathcal{F}_p$-statistic which assume a monochromatic signal over the observation window, our model strictly preserves the time dependence of $\omega(t)$ during the entire $\sim 10$-year observation span. This explicit inclusion of radiation-driven chirping ensures phase coherence for massive or high-frequency binaries that evolve noticeably within the observation window.

To simplify the equations, we define the transformed antenna patterns $F_{+}^{\prime} \equiv F_{+} \cos 2 \psi + F_{\times} \sin 2 \psi$, and $F_{\times}^{\prime} \equiv F_{\times} \cos 2 \psi - F_{+} \sin 2 \psi$. Then we can rewrite the timing residuals $s(t, \hat{\Omega})$ as:
\begin{multline}
s(t, \hat{\Omega}) = \frac{\mathcal{M}^{5/3}}{d_L \omega_p^{1/3}} \big[ -F_{+}^{\prime} (1 + \cos^2 \iota) \sin 2\Phi(t_p) \\
+ 2 F_{\times}^{\prime} \cos \iota \cos 2\Phi(t_p) \big] \\
- \frac{\mathcal{M}^{5/3}}{d_L \omega_e^{1/3}} \big[ -F_{+}^{\prime} (1 + \cos^2 \iota) \sin 2\Phi(t) \\
+ 2 F_{\times}^{\prime} \cos \iota \cos 2\Phi(t) \big]\;.
\end{multline}
where $\omega_p \equiv \omega(t_p)$ and $\omega_e \equiv \omega(t)$ represent the evaluated frequencies at the pulsar and Earth, respectively, separated by thousands of years of evolution due to the light-travel time.

To simplify the expression, we define the following amplitude coefficients and the intrinsic phase shift:
\begin{align}
a &\equiv \frac{\mathcal{M}^{5/3}}{d_L \, \omega_p^{1/3}} \sqrt{F_{+}^{\prime 2} (1 + \cos^2 \iota)^2 + 4 F_{\times}^{\prime 2} \cos^2 \iota}\;, \label{eq:coeff_a} \\
b &\equiv \frac{\mathcal{M}^{5/3}}{d_L \, \omega_e^{1/3}} \sqrt{F_{+}^{\prime 2} (1 + \cos^2 \iota)^2 + 4 F_{\times}^{\prime 2} \cos^2 \iota}\;, \label{eq:coeff_b} \\
\phi_0 &\equiv \arg \left[ -F_{+}^{\prime} (1 + \cos^2 \iota) + i \, 2 F_{\times}^{\prime} \cos \iota \right]\;. \label{eq:phi_0}
\end{align}
With these definitions, the timing residual $s(t, \hat{\Omega})$ collapses into a concise form:
\begin{equation}
s(t, \hat{\Omega}) = a \sin \left( 2 \Phi(t_p) + \phi_0 \right) - b \sin \left( 2 \Phi(t) + \phi_0 \right)\;.
\end{equation}

The distance to the $I$-th pulsar, $L_I$, affects the timing residuals through two distinct mechanisms: the frequency evolution (since the light-travel time dictates the absolute difference between $\omega_p$ and $\omega_e$) and the absolute phase shift. Because typical astronomical measurements of pulsar distances carry uncertainties of $\sim 10\%$ to $20\%$ (often corresponding to hundreds of parsecs), the resulting phase uncertainty spans hundreds or thousands of full GW cycles, effectively randomizing the phase argument. Conversely, this same distance uncertainty results in only a negligible fractional error in the slowly evolving frequency $\omega_p$.

If a search template were strictly parameterized by a single physical distance $L_I$, the minute deviations during parameter space sampling would cause the template's pulsar term to drift rapidly in and out of phase with the data. To construct a more robust search model, we mathematically decouple the macroscopic time-delay from the highly sensitive phase shift. We allow $L_I$ to dictate only the deterministic frequency evolution $\omega_{p,I}$, and introduce an independent, extrinsic phase parameter $\phi_I$ to absorb the sub-cycle phase uncertainty. Under this parameterization, we map the theoretical pulsar phase for the $I$-th pulsar as $\Phi_I(t_p) \rightarrow \Phi_I(t_p) + \phi_I$. The modeled timing residual then assumes a linearly separable form: 
\begin{equation}\label{eq:s_pulsar_term_short}
s_I(t, \hat{\Omega}) = X_I \cos (2 \phi_I) + Y_I \sin (2 \phi_I) + Z_I\;,
\end{equation}
where $X_I = a_I \sin \left( 2 \Phi_I(t_p) + \phi_0 \right)$, $Y_I = a_I \cos \left( 2 \Phi_I(t_p) + \phi_0 \right)$, and $Z_I = -b_I \sin \left( 2 \Phi_I(t) + \phi_0 \right)$.

Crucially, the functions $X_I$, $Y_I$, and $Z_I$ encapsulate the time dependence and global source parameters, and are strictly independent of the pulsar phase $\phi_I$. By intentionally breaking the deterministic link between the geometric distance and the pulsar phase, we treat the $2N_p$ pulsar-specific parameters as $N_p$ distance parameters ($L_I$) that drive the frequency evolution, and $N_p$ extrinsic phase parameters ($\phi_I$). This structural reorganization isolates the nuisance phases into a linear combination, paving the way for the analytical likelihood maximization detailed in the next section.

\section{Methods}\label{sec:Methods}
In this section, we detail the frequentist time-domain search algorithm. We first introduce the PTA data model and the likelihood function. Next, we describe the analytical maximization of the likelihood with respect to the pulsar phases $\phi_I$ via a quartic polynomial. Finally, we outline the joint estimation of both the CGW and red noise parameters using PSO, and define our optimal detection statistic based on the GLRT.

\subsection{PTA Likelihood} 
PTA observations consist of pulse times of arrival (TOA) for a collection of millisecond pulsars. 
In addition to TOAs, PTA data contains pulsar timing models with prescriptions of how pulsar specific deterministic effects including pulsar kinematics, dynamics, as well as astrometry, contribute to TOAs. 
A timing residual is $\delta t={\rm TOA}-\epsilon_0$, where $\epsilon_0=\delta t_{\rm TM}(\hat{\boldsymbol{\theta}}_{\rm TM})$ is the time series prediction of the timing model for a given pulsar, $\delta t_{\rm TM}$, for best-fit timing model parameters $\hat{\theta}_{\rm TM}$.
A vector of timing residuals for all TOAs and all pulsars, where appropriate, is $\boldsymbol{\delta t}$.
As all contemporary full-PTA analyses, we extrapolate the time series predictions of the timing model from $\hat{\boldsymbol{\theta}}_{\rm TM}$ to arbitrary $\boldsymbol{\theta}_{\rm TM}$ using the terms up to linear order in Taylor series, $\boldsymbol{\delta t}_{\rm TM}=\boldsymbol{\epsilon}_0+\boldsymbol{M}\boldsymbol{\epsilon}$. 
Here, $\boldsymbol{M}$ denotes the design matrix representing $\frac{d\delta t_{\rm TM}}{d\theta_{\rm TM}}|_{\hat{\theta}_{\rm TM}}$.
With this, instead of fitting for timing model parameters of pulsar timing models, we fit for linear coefficients of pulsar timing models, $\boldsymbol{\epsilon}$, and $\hat{\boldsymbol{\theta}}_{\rm TM}$ remain as latent fixed parameters.

The post-fit timing residuals are modeled as a linear combination of distinct noise components and deterministic signals,
\begin{equation}
\boldsymbol{\delta t} =
\boldsymbol{M}\boldsymbol{\epsilon} + \boldsymbol{n}_\mathrm{RN} + \boldsymbol{n}_\mathrm{CRN} + \boldsymbol{n}_\mathrm{WN} + \boldsymbol{n}_\mathrm{ECORR} + \boldsymbol{s}\;,\label{eq:residuals}
\end{equation}
where $\boldsymbol{n}_\mathrm{RN}$ represents pulsar-intrinsic low-frequency temporally-correlated ``red'' noise, while $\boldsymbol{n}_\mathrm{CRN}$ denotes red-noise processes common to all pulsars, such as a stochastic GWB. 
Temporally-uncorrelated white noise term $\boldsymbol{n}_\mathrm{WN}$ includes TOA uncertainty corrections, error factor (EFAC), error added in quadrature (EQUAD) parameters. Epoch-correlated noise, ECORR, is white noise which is identical for TOAs in one observing epoch, but uncorrelated between any two observing epochs. 
The vector $\boldsymbol{s}$ corresponds to the deterministic CGW signal from an individual SMBHB, as derived in Eq.~\eqref{eq:s_pulsar_term_short}.
Before proceeding further, we compactify Equation~\ref{eq:residuals} as $\boldsymbol{\delta t}=\boldsymbol{T}\boldsymbol{b}+\boldsymbol{n}_\mathrm{WN}+\boldsymbol{s}$, where $\boldsymbol{T}=[\boldsymbol{M}\;\boldsymbol{F}\;\boldsymbol{J}]$. Following the standard linear-basis formulation of PTA likelihood \cite{lentati_hyper-efficient_2013, van_haasteren_new_2014}, $\boldsymbol{F}$ denotes the matrix block containing the Fourier design matrix used for the red-noise and common-red-noise basis functions, $\boldsymbol{J}$ denotes the matrix block containing the ECORR epoch design matrix, and $\boldsymbol{b}$ collects the linear coefficients associated with the timing-model, Fourier, and ECORR bases.
The corresponding coefficient covariance is denoted by $\boldsymbol{B}$; after marginalizing over $\boldsymbol{b}$, the total covariance is $\boldsymbol{C}=\boldsymbol{N}+\boldsymbol{T}\boldsymbol{B}\boldsymbol{T}^{\intercal}$, where $\boldsymbol{N}$ is the white noise covariance matrix.

Because the present study includes only mutually uncorrelated pulsar-intrinsic red noise and no GWB, we model the timing-residual power spectral density of pulsar $I$ directly as
\begin{equation}
P_{{\rm RN},I}(f)=\frac{A_{{\rm RN},I}^{2}}{12\pi^{2}}
\left(\frac{f}{f_{\rm yr}}\right)^{-\gamma_{{\rm RN},I}}f_{\rm yr}^{-3},
\label{eq:rn_psd}
\end{equation}
where $A_{{\rm RN},I}$ and $\gamma_{{\rm RN},I}$ are the intrinsic red-noise amplitude and spectral index of that pulsar, respectively, and $f_{\rm yr}=1/\mathrm{yr}$.

Let $\boldsymbol{q}$ denote the full set of model parameters, including both deterministic signal parameters and noise hyperparameters. 
With the covariance defined above, the PTA likelihood marginalized over coefficients $\boldsymbol{b}$ is a multivariate Gaussian distribution:
\begin{equation}
\mathcal{L}(\boldsymbol{\delta t} | \boldsymbol{q}) = 
\frac{
\exp\left[ 
-\frac{1}{2} (\boldsymbol{\delta t} - \boldsymbol{s})^\intercal \boldsymbol{C}^{-1} (\boldsymbol{\delta t} - \boldsymbol{s}) 
\right]
}{
\sqrt{ \det(2\pi \boldsymbol{C}) }
} \;. \label{eq:likelihood}
\end{equation}

The key difference of our proposed approach from the traditional $\mathcal{F}$~statistic is described below.
Taking the natural logarithm of Eq.~\eqref{eq:likelihood}, we obtain
\begin{equation}
\ln \mathcal{L} = 
-\frac{1}{2} (\boldsymbol{\delta t} - \boldsymbol{s})^\intercal \boldsymbol{C}^{-1} (\boldsymbol{\delta t} - \boldsymbol{s})
-\frac{1}{2} \ln \det(2\pi \boldsymbol{C}) \;.
\end{equation}

Rather than serving directly as the detection statistic, this full log-likelihood acts as the objective function for our joint parameter estimation. It is crucial to distinguish this optimization approach from traditional $\mathcal{F}$-statistics.
Standard frequentist methods are based on the log-likelihood ratio $\ln \Lambda$, which is explicitly defined in terms of the signal hypothesis ($H_1$) and the noise-only hypothesis ($H_0$, where $\boldsymbol{s}=\boldsymbol{0}$):
\begin{equation}
\ln \Lambda \equiv \ln \mathcal{L}_{H_1} - \ln \mathcal{L}_{H_0} = 
\boldsymbol{\delta t}^\intercal \boldsymbol{C}^{-1} \boldsymbol{s} - \frac{1}{2} \boldsymbol{s}^\intercal \boldsymbol{C}^{-1} \boldsymbol{s} \;.
\end{equation}
Consequently, the full log-likelihood under the signal hypothesis can be decomposed as:
\begin{equation}
\ln \mathcal{L}_{H_1} = \ln \Lambda 
- \frac{1}{2} \boldsymbol{\delta t}^{\intercal} \boldsymbol{C}^{-1} \boldsymbol{\delta t} 
- \frac{1}{2} \ln \det(2\pi \boldsymbol{C}) \;.
\end{equation}

Traditional frequentist $\mathcal{F}$-statistic searches based solely on maximizing $\ln \Lambda$ (such as the $\mathcal{F}_e$- and $\mathcal{F}_p$-statistics) require fixing the noise covariance $\boldsymbol{C}$ beforehand to maximize $\ln \Lambda$. Thus, these techniques do not treat the noise model concurrently with the signal, and fitting the model to data is essentially performed twice. First, for the noise. Second, for the signal. 
This separation makes searches based on the $\mathcal{F}$ statistic highly susceptible to noise-to-signal leakage. 
In this work, we propose to maximize the complete $\ln \mathcal{L}$, evaluating the noise covariance $\boldsymbol{C}$ alongside the signal $\boldsymbol{s}$, enabling a rigorous joint estimation that breaks the degeneracy between intrinsic red noise and continuous waves. A related full-likelihood viewpoint was also noted in recent $\mathcal{F}$-statistic work~\cite{furusawa_resolving_2025}.

\subsection{Analytical elimination of the pulsar phase}
\label{sec:pulsar_phase_maximization}
For a PTA with $N_p$ pulsars, the total deterministic signal is the concatenation of the contributions from individual pulsars, with the timing residual for the $I$-th pulsar given in Eq.~\eqref{eq:s_pulsar_term_short}. The log-likelihood function, defined in Eq.~\eqref{eq:likelihood}, can be expressed as the sum of single-pulsar likelihoods when the noise is assumed to be uncorrelated between pulsars. Hence, the maximization with respect to each pulsar’s phase parameter $\phi_I$ can be performed independently.

When all parameters except $\phi_I$ are fixed, it has been found that the log-likelihood in Eq.~\eqref{eq:likelihood} is a quadratic function of $s_I$ and can be maximized analytically with respect to $\phi_I$ \cite{wang_coherent_2015}. The dependence on $\phi_I$ enters only through Eq.~\eqref{eq:s_pulsar_term_short}, so the maximization reduces to a one-dimensional problem for each pulsar.

Defining the noise-weighted inner product as
\begin{equation}
    (\boldsymbol{x}, \boldsymbol{y}) \equiv \boldsymbol{x}^{\intercal} \boldsymbol{C}_I^{-1} \boldsymbol{y} \;,
\label{eq:inner_product}
\end{equation}
the log-likelihood ratio for pulsar $I$ can be written as
\begin{equation}
    \ln \Lambda_I = (\boldsymbol{r}_I, \boldsymbol{s}_I) - \frac{1}{2} (\boldsymbol{s}_I, \boldsymbol{s}_I)\;,
\label{eq:loglike_pulsar}
\end{equation}
where $\boldsymbol{r}_I$ is the vector of observed timing residuals and $\boldsymbol{C}_I$ is the noise covariance matrix for the $I$-th pulsar. Let $\boldsymbol{X}_I$, $\boldsymbol{Y}_I$, and $\boldsymbol{Z}_I$ denote the corresponding discretely sampled vectors of the time-dependent basis functions defined previously. Substituting Eq.~\eqref{eq:s_pulsar_term_short} into Eq.~\eqref{eq:loglike_pulsar}, the single-pulsar log-likelihood ratio reads
\begin{equation}
\begin{split}
\ln \Lambda_I &=
  (\boldsymbol{r}_I, \boldsymbol{X}_I) \cos 2\phi_I
+ (\boldsymbol{r}_I, \boldsymbol{Y}_I) \sin 2\phi_I
+ (\boldsymbol{r}_I, \boldsymbol{Z}_I)  \\
&\quad
- \frac{1}{2} \Big[
   (\boldsymbol{X}_I, \boldsymbol{X}_I) \cos^2 2\phi_I
 + (\boldsymbol{Y}_I, \boldsymbol{Y}_I) \sin^2 2\phi_I  \\
&\qquad
 + 2 (\boldsymbol{X}_I, \boldsymbol{Y}_I) \sin 2\phi_I \cos 2\phi_I
 + 2 (\boldsymbol{X}_I, \boldsymbol{Z}_I) \cos 2\phi_I  \\
&\qquad
 + 2 (\boldsymbol{Y}_I, \boldsymbol{Z}_I) \sin 2\phi_I
 +   (\boldsymbol{Z}_I, \boldsymbol{Z}_I)
 \Big]\;.
\end{split}
\label{eq:loglike_trig}
\end{equation}

The analytic maximization of Eq.~\eqref{eq:loglike_trig} with respect to $\phi_I$ can be achieved by introducing the variable $y = \cos 2\phi_I$, which transforms the problem into solving a quartic polynomial of the form
\begin{equation}
a y^4 + b y^3 + c y^2 + d y + e = 0 \;,
\label{eq:quartic_cos}
\end{equation}
where the coefficients $a$, $b$, $c$, $d$, and $e$ are algebraic functions of the inner products defined in Eq.~\eqref{eq:loglike_trig}:
\begin{subequations}
\label{eq:abcd_from_c1234}
\begin{align}
a &= [(\boldsymbol{X}_I,\boldsymbol{X}_I)-(\boldsymbol{Y}_I,\boldsymbol{Y}_I)]^2 + 4(\boldsymbol{X}_I,\boldsymbol{Y}_I)^2\;, \\[2pt]
b &= 2[(\boldsymbol{X}_I,\boldsymbol{X}_I)-(\boldsymbol{Y}_I,\boldsymbol{Y}_I)][(\boldsymbol{X}_I,\boldsymbol{Z}_I)-(\boldsymbol{r}_I,\boldsymbol{X}_I)] \nonumber\\
  &\quad +\,4(\boldsymbol{X}_I,\boldsymbol{Y}_I)[(\boldsymbol{Y}_I,\boldsymbol{Z}_I)-(\boldsymbol{r}_I,\boldsymbol{Y}_I)]\;, \\[2pt]
c &= [(\boldsymbol{X}_I,\boldsymbol{Z}_I)-(\boldsymbol{r}_I,\boldsymbol{X}_I)]^2 + [(\boldsymbol{r}_I,\boldsymbol{Y}_I)-(\boldsymbol{Y}_I,\boldsymbol{Z}_I)]^2 \nonumber\\
  &\quad -[(\boldsymbol{X}_I,\boldsymbol{X}_I)-(\boldsymbol{Y}_I,\boldsymbol{Y}_I)]^2 - 4(\boldsymbol{X}_I,\boldsymbol{Y}_I)^2\;, \\[2pt]
d &= -2[(\boldsymbol{X}_I,\boldsymbol{X}_I)-(\boldsymbol{Y}_I,\boldsymbol{Y}_I)][(\boldsymbol{X}_I,\boldsymbol{Z}_I)-(\boldsymbol{r}_I,\boldsymbol{X}_I)] \nonumber\\
  &\quad -\,2(\boldsymbol{X}_I,\boldsymbol{Y}_I)[(\boldsymbol{Y}_I,\boldsymbol{Z}_I)-(\boldsymbol{r}_I,\boldsymbol{Y}_I)]\;, \\[2pt]
e &= (\boldsymbol{X}_I,\boldsymbol{Y}_I)^2 - [(\boldsymbol{X}_I,\boldsymbol{Z}_I)-(\boldsymbol{r}_I,\boldsymbol{X}_I)]^2 \;.
\end{align}
\end{subequations}

This quartic equation admits up to four real roots, corresponding to up to four candidate values for $2\phi_I$. 
For each real root, as well as the boundary values $\phi_I \in [0, \pi)$ (for which $2\phi_I \in [0, 2\pi)$), the value of $\ln  \Lambda_I$ is evaluated directly, and the $\phi_I$ that maximizes the log likelihood ratio is selected.

This procedure is performed independently for each pulsar, effectively removing the pulsar-phase parameters from the global search space. Following Ref.~\cite{wang_coherent_2015}, this analytic maximization reduces the dimensionality of the optimization problem from $2N_p+8$ to $N_p+8$, substantially improving the computational efficiency of the search without numerically sampling the pulsar phases.

\begin{figure}[htb]
    \centering
    \includegraphics[width=\linewidth]{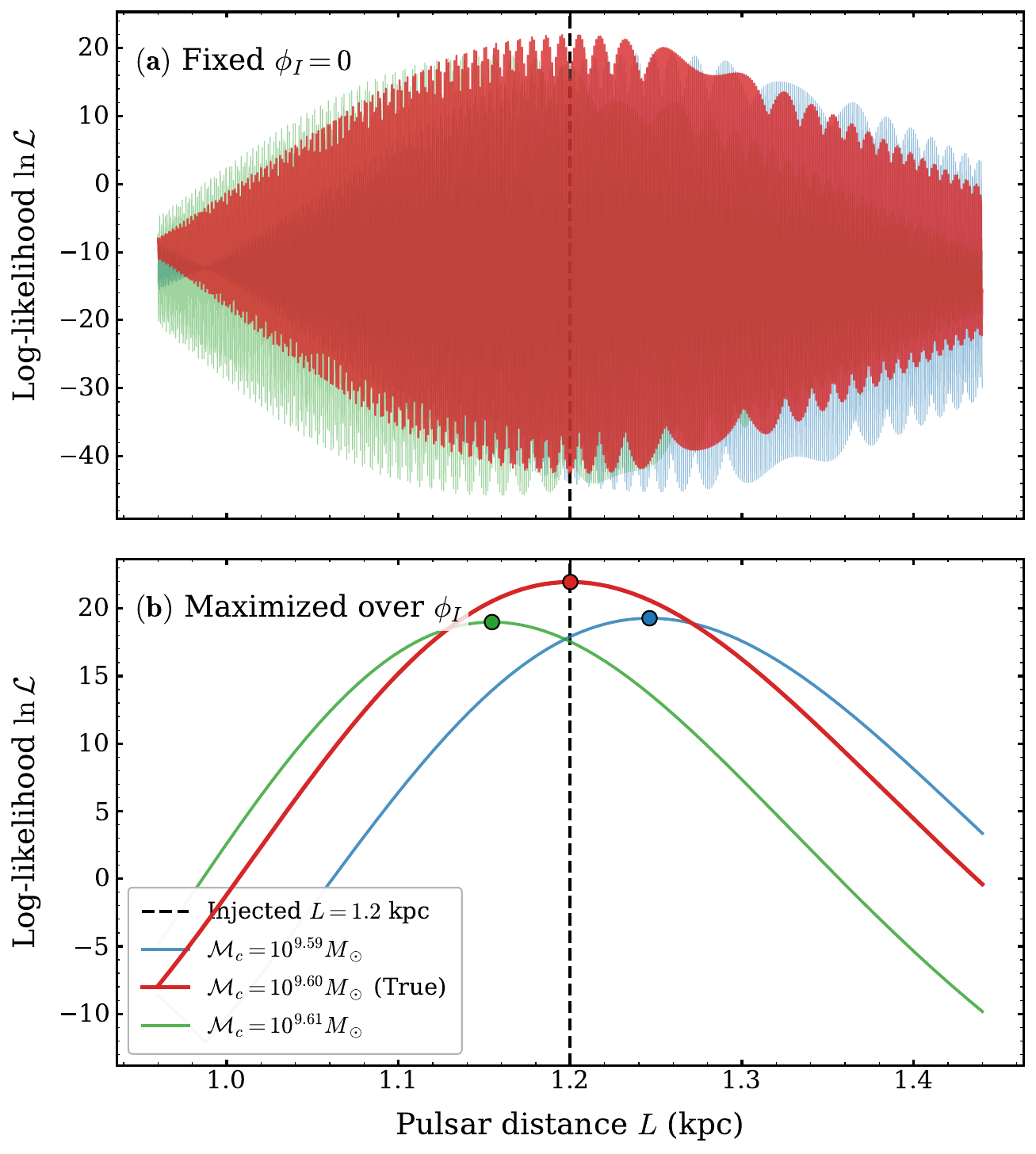}
    \caption{Log-likelihood $\ln \mathcal{L}$ as a function of trial pulsar distance $L$ for a simulated rapidly-evolving CGW signal. The signal parameters are: $\log_{10}(h) = -14.0$, $\log_{10}(f_{\mathrm{gw}}/\mathrm{Hz}) = -7.3$, $\log_{10}(\mathcal{M}_c/M_{\odot}) = 9.60$, $L = 1.2$~kpc, $\cos\theta = 0.12$, $\phi = 3.2$, $\cos\iota = 0.3$, $\psi = 1.2$, and $\phi_I = 1.6$. (a) Likelihood evaluated with a fixed initial phase $\phi_I = 0$, exhibiting phase oscillations. (b) Likelihood analytically maximized over $\phi_I$, revealing the smooth envelope. The colored curves represent recovery templates with varying chirp mass assumptions ($\log_{10}\mathcal{M}_c \in \{9.59, 9.60, 9.61\}$), where the injected value (red) correctly recovers the peak at the dashed vertical line ($L = 1.2$~kpc).}\label{fig:phase_maximization_demo}
\end{figure}

To visually demonstrate the effect of this analytical elimination, Figure~\ref{fig:phase_maximization_demo} illustrates the single-pulsar log-likelihood ratio as a function of the trial pulsar distance $L_I$. When $\phi_I$ is fixed to $0$, the likelihood surface is dominated by high-frequency interference fringes caused by the extreme sensitivity of the light-travel time to microscopic distance variations. By analytically maximizing over $\phi_I$, we remove these phase fringes from the numerical search over $L_I$. This transformation reduces the dimensionality of the search space and makes the remaining optimization more tractable for stochastic sampling algorithms.

\subsection{Joint Estimation via Particle Swarm Optimization} \label{sebsec:pso}

With the $N_p$ highly uncertain pulsar phase parameters analytically eliminated from the search space, the remaining parameters are treated as intrinsic parameters that must be optimized numerically. This intrinsic parameter space is exceptionally high-dimensional, comprising the global CGW parameters $(\mathcal{M}, \omega_0, \Phi_0, \iota, \psi, d_L, \alpha, \delta)$, the pulsar distances $L_I$, and critically, the individual red noise amplitude and spectral index parameters for every pulsar in the array. 

Performing a simultaneous search over this joint signal-and-noise parameter space is the cornerstone of our method to prevent noise-to-signal leakage. However, the joint likelihood landscape is notoriously complex and highly multimodal, driven by strong covariances between the low-frequency deterministic CGW signal and the stochastic red noise realization. To robustly locate the global maximum of the full log-likelihood $\ln \mathcal{L}$ without being trapped in local maxima (which often correspond to biased, signal-leaked fits), we employ the PSO algorithm. PSO has been extensively and successfully applied in gravitational-wave data analysis for PTA~\citep{2017PhRvL.118o1104W,PhysRevD.106.023016,2025PhRvD.111h3030T}, ground-based~\cite{Wang_2010,weerathunga_performance_2017} and space-based~\cite{zhang_resolving_2021,2025PhRvD.111d3026F} interferometers. In what follows, we provide a brief overview of the PSO mechanism and describe the specific configuration adopted to yield the Maximum Likelihood Estimators (MLE) for our frequentist statistic.

PSO is an iterative, population-based stochastic optimization method designed to locate the global maximum of a fitness function $f(x)$, $x \in \mathbb{R}^N$, over a compact search domain $D \subset \mathbb{R}^N$. In our framework, the fitness function strictly corresponds to the dynamically evaluated log-likelihood $\ln \mathcal{L}$ (see Eq.~\ref{eq:likelihood}). During each iteration, the algorithm evaluates the fitness function at a set of sample locations (particles), updates their velocities based on gathered information, and iteratively guides the swarm toward regions of higher likelihood. The swarm size remains fixed throughout the process.

Let $x_{i,j}[k]$ and $v_{i,j}[k]$ denote the position and velocity of the $I$-th particle at iteration $k$, where $i \in \{1, 2, \ldots, N_{\rm part}\}$ and $j$ indexes the parameter component. The particle is first assigned a trial velocity, that velocity is clipped component-wise, and the clipped velocity is then used to update the position. We use a symmetric velocity cap in each parameter direction:
\begin{subequations}
\begin{align}
v^{\rm raw}_{i,j}[k+1] &= w[k]v_{i,j}[k]
+ c_1 r_{1,i,j}[k]\bigl(p_{i,j}[k]-x_{i,j}[k]\bigr) \notag \\
&\quad + c_2 r_{2,i,j}[k]\bigl(l_{i,j}[k]-x_{i,j}[k]\bigr), \\
\operatorname{clip}(u;v_{\max,j}) &= \min\!\left\{v_{\max,j},\max\!\left\{-v_{\max,j},u\right\}\right\}, \\
v_{i,j}[k+1] &= \operatorname{clip}\!\left(v^{\rm raw}_{i,j}[k+1];v_{\max,j}\right), \\
x_{i,j}[k+1] &= x_{i,j}[k] + v_{i,j}[k+1] \,,
\end{align}
\end{subequations}
where $v_{\rm max}$ is the upper limit on the step size to prevent chaotic divergence, $w[k]$ is the inertia weight that decreases with the iteration number, and $c_1$, $c_2$ are acceleration coefficients. The random variables $r_1$ and $r_2$ are uniformly distributed in $[0, 1]$. Each particle maintains a memory of its own previous position corresponding to the highest likelihood, $p_i[k]$ (its \emph{personal best}), while $l_i[k]$ denotes the position yielding the highest likelihood found within its local topological neighborhood.

Each term in the velocity update serves a distinct search function: the inertia term promotes motion along the current trajectory, aiding in broad exploration and helping the swarm plow through local likelihood ridges; the cognitive term attracts the particle toward its own historical success; and the social term pulls it toward the success of its neighbors. To delay premature convergence and preserve swarm diversity across the massive parameter space, we adopt a local ring topology, where particles are arranged cyclically and share information only with their immediate nearest neighbors, rather than instantaneously collapsing toward a single global best $g[k]$.

We find that the PSO method performs most efficiently when the search domain $D$ is hypercubic, with each parameter component $x_j$ confined to a physically motivated prior boundary $[a_j, b_j]$. The initial positions and velocities of all particles are drawn randomly from these intervals. To enforce these strict boundaries during later iterations, we apply a ``let-them-fly'' boundary condition: any particle attempting to explore outside the valid parameter domain $D$ is instantly assigned a heavily penalized fitness value of $-\infty$, forcing the swarm to refocus its search internally. 

The optimization process terminates after a fixed number of iterations. The final global best position represents the MLE required to evaluate the detection statistic $\mathcal{T}$ to be defined in Sec.~\ref{subsec:threshold_method}. Unlike MCMC methods, which require extensive computational time to overcome burn-in and generate uncorrelated posterior samples, PSO approaches the likelihood peak directly. Although a single PSO run does not mathematically guarantee absolute convergence to the exact global maximum, the success probability $P_{\rm success}$ is substantially amplified by executing the optimization iteratively with independent random seeds. For $N_{\rm runs}$ independent runs, the probability of at least one successful convergence scales as $1 - (1 - P_{\rm success})^{N_{\rm runs}}$. 

In our implementation, we perform $N_{\rm runs} = 4$ independent PSO runs in parallel to ensure the utmost robustness against the complex degeneracies of the joint noise-signal space. The internal configuration closely mirrors established standards: we deploy $N_{\text{part}} = 40$ particles over $N_{\text{iter}} = 2000$ iterations. The acceleration coefficients are symmetrically balanced at $c_1 = c_2 = 2.0$, with a topological neighborhood size of $m = 3$. The maximum velocity is initialized as $v_{\max} = (b-a)/2$ and aggressively restricted to $(b-a)/5$ in subsequent stages to allow the swarm to fine-tune to the likelihood maximum. Finally, the inertia weight dictates a smooth transition from global exploration to localized exploitation via a linear decay:
\begin{equation}
w[k] = 0.9 - 0.5\,\frac{k}{N_{\text{iter}} - 1}\;.
\end{equation}

\subsection{Detection Statistic and Threshold}
\label{subsec:threshold_method}
To quantify the significance of a potential detection, we employ a detection statistic, denoted as $\mathcal{T}$, constructed based on the GLRT principle \cite{kay1998fundamentals}. Instead of adopting the hybrid ``noise-marginalized'' approach recently used in some PTA data releases---which inconsistently integrates a frequentist statistic over a Bayesian noise posterior---we strictly adhere to a frequentist profile likelihood approach. Here, the nuisance parameters (i.e., the intrinsic red noise parameters for each pulsar) are treated as fully free parameters and are simultaneously maximized under binary hypotheses.

The detection statistic is defined as the difference between the maximum log-likelihood values obtained under the signal hypothesis ($H_1$) and the null hypothesis ($H_0$):
\begin{equation}
\mathcal{T} = \ln \mathcal{L}_{H_1}(\hat{\boldsymbol{\theta}}_{\mathrm{gw}}, \hat{\boldsymbol{\theta}}_{\mathrm{noise}}) - \ln \mathcal{L}_{H_0}(\hat{\boldsymbol{\theta}}_{\mathrm{noise}}') \;,
\label{eq:detection_statistic}
\end{equation}
where $\hat{\boldsymbol{\theta}}_{\mathrm{gw}}$ denotes the maximum-likelihood estimators (MLEs) for the GW signal parameters (e.g., frequency, sky location, and chirp mass), while $\hat{\boldsymbol{\theta}}_{\mathrm{noise}}$ and $\hat{\boldsymbol{\theta}}_{\mathrm{noise}}'$ denote the MLEs for the red-noise parameters under $H_1$ and $H_0$, respectively. Because the covariance is profiled independently under the signal and noise-only hypotheses, the statistic compares the best signal-plus-noise explanation with the best noise-only explanation without forcing both hypotheses to share a single covariance matrix. This suppresses the systematic absorption of CGW power into the red-noise model, while still allowing finite-realization stochastic degeneracies between weak red noise and the deterministic template.

The two hypotheses are defined as follows:
\begin{itemize}
    \item \textbf{Signal Hypothesis ($H_1$):} The data contains both a deterministic CGW signal and stochastic noise. The log-likelihood $\ln \mathcal{L}_{H_1}$ is maximized over the full parameter space, including the extrinsic and intrinsic GW parameters as well as the red noise parameters (red noise amplitude and spectral index) for each pulsar. This joint maximization is performed using the PSO algorithm to search for the global optimum.
    \item \textbf{Null Hypothesis ($H_0$):} The data contains only stochastic noise. The log-likelihood $\ln \mathcal{L}_{H_0}$ is maximized solely over the red noise parameters for each pulsar. This step provides the baseline likelihood for the fully optimized noise-only model.
\end{itemize}

The theoretical motivation for this formulation follows the standard GLRT treatment of binary hypothesis tests with unknown parameters \cite{kay1998fundamentals}. For simple hypotheses, the Neyman--Pearson lemma identifies the likelihood-ratio test as the most powerful test at fixed false-alarm probability \cite{Neyman_1933}. For composite hypotheses, no uniformly optimal statistic exists in general. The GLRT replaces the unknown parameters under each hypothesis with their maximum-likelihood estimates and can have asymptotic optimality properties under regularity conditions. In the present problem, this construction differs structurally from $\mathcal{F}_e$- and $\mathcal{F}_p$-type statistics because the noise covariance is not held fixed or imported from an external analysis: it is re-estimated under both $H_1$ and $H_0$. Consequently, when signal and red noise are covariant, a fixed or externally inferred noise model need not represent the best noise description under either competing hypothesis, which can affect detection performance and parameter recovery.

Establishing a detection threshold requires determining the probability distribution of $\mathcal{T}$ under the null hypothesis ($H_0$). Standard statistical theory (Wilks' theorem) predicts that for nested models, the likelihood ratio statistic asymptotically follows a $\chi^2$ distribution with degrees of freedom equal to the difference in the number of free parameters~\cite{Wilks_1938}. However, this theorem holds only when the parameters of the fuller model are mathematically well-defined under the null hypothesis~\cite{davies_hypothesis_1987}.

Unlike the standard $\mathcal{F}_e$-statistic, which typically assumes a fixed-frequency monochromatic signal and adheres to a $\chi^2$ distribution under fixed noise assumptions, our method incorporates highly non-linear parameters---such as the signal frequency, sky location, and frequency evolution (chirp mass)---that simply do not exist under $H_0$. Furthermore, the global maximization of the likelihood over this vast, undefined parameter space introduces a massive ``look-elsewhere'' effect (also known as the trials factor). Consequently, the detection statistic $\mathcal{T}$ no longer follows a standard $\chi^2$ distribution. Instead, the global supremum of the likelihood over the parameter space adopts an analytically intractable extreme value distribution  (EVD) \cite{leadbetter_extremes_1983}, characterized by a significantly heavier tail at high values.

Since the effective number of independent spatial and frequency trials, as well as the precise shape of the EVD tail, are analytically intractable due to the complex correlations in PTA red noise, we must estimate the null distribution empirically. We define the false alarm probability (FAP), denoted as $\alpha$, as the probability that the detection statistic exceeds a certain threshold $\mathcal{T}_{\mathrm{th}}$ under $H_0$:
\begin{equation}
\alpha = P(\mathcal{T} > \mathcal{T}_{\mathrm{th}} | H_0) \;.
\label{eq:fap_def}
\end{equation}

To determine $\mathcal{T}_{\mathrm{th}}$ for a target FAP, we employ an empirical Monte Carlo approach. We generate a large set of independent noise-only realizations ($N_{\mathrm{noise}}$) consistent with the noise properties of the dataset. For each realization, we perform the full PSO joint optimization under both $H_1$ and $H_0$ to calculate the empirical $\mathcal{T}$. The threshold $\mathcal{T}_{\mathrm{th}}(\alpha)$ is then defined as the $(1-\alpha)$-quantile of this empirical distribution. While running these simulations requires upfront computational resources, it completely avoids the philosophical inconsistencies of hybrid Bayesian-frequentist marginalization and establishes an empirically calibrated frequentist threshold tailored to the specific spatial rigidity and noise profile of the PTA.

Finally, the detection performance is quantified by the detection probability (or efficiency) $P_d$, defined as the probability that the statistic exceeds the established threshold when a signal is present:
\begin{equation}
P_d(\rho) = P(\mathcal{T} > \mathcal{T}_{\mathrm{th}}(\alpha) | H_1(\rho)) \;,
\label{eq:pd_def}
\end{equation}
where $\rho$ denotes the SNR of the injected signal. For the $I$-th pulsar, the individual optimal SNR is defined by $\rho_I^2=(\boldsymbol{s}_I,\boldsymbol{s}_I)$, where $(\cdot,\cdot)$ is the noise-weighted inner product defined in Eq.~\eqref{eq:inner_product}. The total array SNR used below is $\rho_{\rm tot}=\left(\sum_{I=1}^{N_{p}}\rho_I^2\right)^{1/2}$.

\section{Results}\label{sec:SimResults}
In the following, we present the main results of our CGW search method. This section is organized as follows. We first describe the simulation setup used to generate realistic PTA datasets with injected CW signals. Then, we discuss the detection performance of the algorithm and the statistical benefits of array scaling. Subsequently, we evaluate the parameter estimation accuracy, demonstrating how our joint estimation framework effectively disentangles red noise from the deterministic signal. We also highlight the critical importance of incorporating time-domain frequency evolution. Finally, we analyze the computational cost and runtime behavior of the proposed algorithm.

\subsection{Simulation} \label{sebsec:sim}
The simulated pulsar timing data used in this work were generated with the \texttt{fakepta} package\footnote{\url{https://github.com/mfalxa/fakepta}},
an open-source tool designed for realistic PTA simulations.
It provides an interface to create synthetic datasets that are directly compatible with the \texttt{ENTERPRISE}~\cite{Ellis_2019_enterprise} and \texttt{enterprise\_extensions}~\cite{enterpriseextensions} frameworks.
This package allows users to specify the number of pulsars, their sky positions, observation cadence, total observing timespan, and various noise components (white and red timing noise).
It also supports the injection of deterministic signals, such as CGWs from supermassive black hole binaries.
The generated datasets are stored as Python \texttt{pickle} objects, which can be directly loaded into \texttt{ENTERPRISE} for subsequent likelihood evaluation and parameter inference.

In our simulations, we generated two distinct PTA configurations to investigate the impact of array size on detection performance:
\begin{itemize}
    \item \textbf{Dataset A (20 Pulsars):} A baseline array comprising 20 pulsars.
    \item \textbf{Dataset B (30 Pulsars):} An extended array comprising 30 pulsars.
\end{itemize}
Both datasets span a total observation time of 10 years and contain between 187 and 210 TOA measurements per pulsar.
The TOAs were initially generated as 261 evenly spaced samples per pulsar with a two-week cadence; subsequently, a random subset of points was removed to introduce irregular sampling.
The white noise for each pulsar was modeled using the EFAC, ECORR, and EQUAD parameters, while the red noise was characterized by a distinct power-law spectrum for each pulsar.

To systematically evaluate the detection performance across different signal strengths, we generated four distinct sub-datasets for each array configuration by adjusting the strain amplitude of the injected CGW signal.
For Dataset A (20 pulsars), the strain amplitudes were set to $\log_{10} h \in \{-20.00, -14.65, -14.35, -14.17\}$.
For Dataset B (30 pulsars), to maintain comparable total SNRs, the amplitudes were adjusted to $\log_{10} h \in \{-20.00, -14.75, -14.48, -14.30\}$.
These values correspond to total array SNRs of approximately $\rho_{\mathrm{tot}} \approx 0, 10, 20$, and $30$, respectively, allowing for a direct comparison of detection statistics under equivalent signal power but different spatial sampling densities.

For all datasets with injected signals (i.e., $\rho_{\mathrm{tot}} > 0$), the pulsar-term phase $\phi_I$ for all pulsars was set to zero.
At this stage, no GWB was injected, as the primary objective is to perform an initial validation of the proposed method's ability to isolate a single CGW against intrinsic pulsar noises.
The injected parameters for the simulated CGW signal are summarized in Table~\ref{tab:cgw_params}.
The specific noise properties and the individual SNR contributions for the pulsars in both datasets are detailed in Appendix~\ref{app:pulsar_params}.

\begin{table}[htbp]
\centering
\caption{Parameters for the simulated CGW signal and their corresponding search ranges used in the PSO optimization.
The strain amplitude $\log_{10} h$ takes four distinct values for each dataset configuration (20 and 30 pulsars), corresponding to SNRs of approximately $0, 10, 20$, and $30$.
}
\label{tab:cgw_params}
\renewcommand{\arraystretch}{1.4}
\begin{tabular}{lcc}
\hline
\hline
\textbf{Parameter} & \textbf{True Value} & \textbf{Search Range} \\ \hline  
& & \\ [-12pt]
$\log_{10} h$ (20 PSRs) & \small \shortstack{$(-20.00, -14.65,$\\$-14.35, -14.17)$} & $[-20,\,-10]$ \\ [1ex]
$\log_{10} h$ (30 PSRs) & \small \shortstack{$(-20.00, -14.75,$\\$-14.48, -14.30)$} & $[-20,\,-10]$ \\ [1ex]
$\cos\theta$ & $0.12$ & $[-1,\,1]$ \\
$\phi~(\mathrm{rad})$ & $3.2$ & $[0,\,2\pi]$ \\
$\cos\iota$ & $0.3$ & $[-1,\,1]$ \\
$\Phi_0~(\mathrm{rad})$ & $1.6$ & $[0,\,2\pi]$ \\
$\psi~(\mathrm{rad})$ & $1.2$ & $[0,\,2\pi]$ \\
$\log_{10} \mathcal{M}_c~(M_\odot)$ & $9.2$ & $[8,\,10]$ \\
$\log_{10} f_{\mathrm{gw}}~(\mathrm{Hz})$ & $-8$ & $[-9,\,-7]$ \\
\hline
\hline
\end{tabular}
\end{table}

The distribution of individual SNRs varies significantly across the array due to differences in noise properties and geometric alignment with the source, as detailed in Appendix~\ref{app:pulsar_params}.

\begin{figure}[htbp]
\centering
\includegraphics[width=1.0\columnwidth]{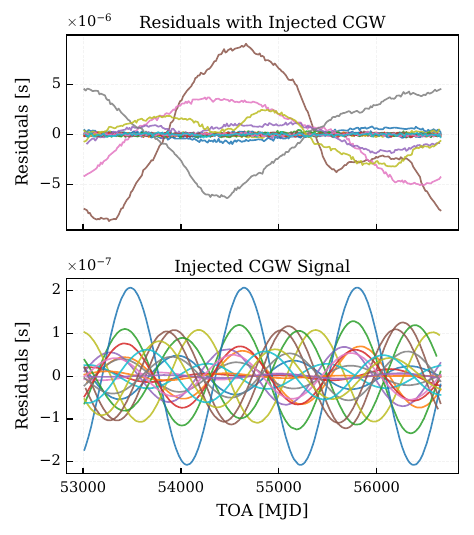}
\caption{Time-domain representation of the simulated data for a representative realization (Dataset A, $\rho_{\mathrm{tot}} \approx 30$). The top panel shows the total simulated timing residuals for all 20 pulsars, and the bottom panel shows the isolated CGW signal.
}
\label{fig:mock_residuals}
\end{figure}

To statistically characterize the detection performance, we generated an ensemble of 50 independent noise realizations for each signal strength scenario. In these Monte Carlo simulations, the intrinsic properties of the pulsars and the CGW signal were held fixed; only the stochastic realization of the white and red noise processes was varied. Figure~\ref{fig:mock_residuals} illustrates the time-domain timing residuals for the 20-pulsar array in a representative realization ($\rho_{\mathrm{tot}} \approx 30$). 
The top panel displays the total simulated residuals. Consistent with the injected parameters listed in Appendix~\ref{app:pulsar_params}, five pulsars with strong injected red noise ($\log_{10} A_{\mathrm{RN}} > -14$) exhibit prominent large-amplitude fluctuations ($\sim \mathcal{O}(10^{-5})\,\mathrm{s}$), while the remaining pulsars ($\log_{10} A_{\mathrm{RN}} \lesssim -15$) possess much quieter profiles. 
The bottom panel isolates the injected deterministic CGW signature, whose amplitude ($\sim \mathcal{O}(10^{-7})\,\mathrm{s}$) is approximately two orders of magnitude smaller than the dominant red noise background.

\subsection{Detection Performance and Array Scaling}
\label{subsec:det_perf}
We evaluated the performance of the GLRT framework defined in Sec.~\ref{subsec:threshold_method} using 50 independent noise realizations for each injected signal strength. For each array, the 50 realizations with $\rho_{\rm tot}=0$ define the empirical null distribution. The array-specific threshold is its $(1-\alpha)$-quantile, with $\alpha=3/50\simeq0.06$, and the resulting threshold is then applied to every ensemble with $\rho_{\rm tot}>0$ for that array.

\begin{figure}[h]
\centering
\includegraphics[width=0.98\columnwidth]{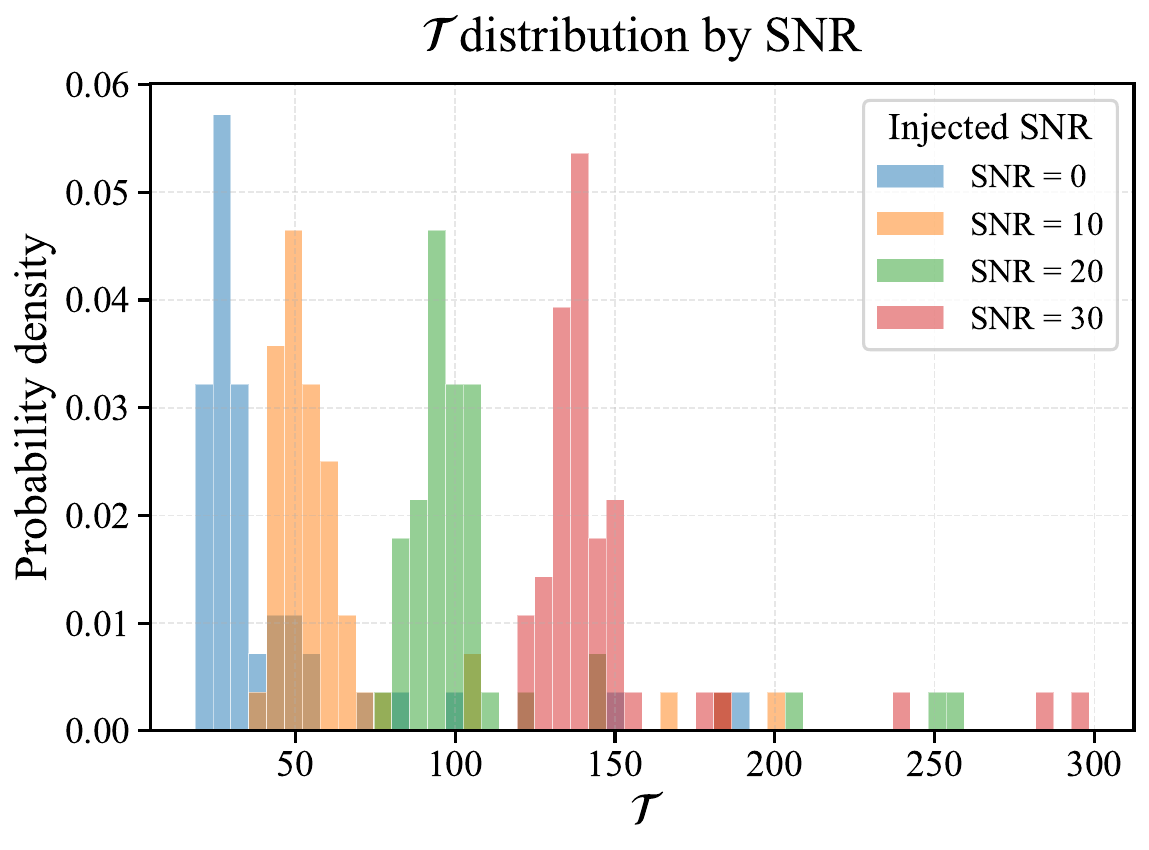}
\caption{
Distributions of the detection statistic $\mathcal{T}$ for four different signal strengths: SNR $\approx$ 0 (blue), 10 (orange), 20 (green), and 30 (red).
Each histogram is constructed from 50 independent realizations using the 20-pulsar array (Dataset A).}
\label{fig:snr_distribution}
\end{figure}

Figure~\ref{fig:snr_distribution} displays the histograms of the recovered detection statistic $\mathcal{T}$ for the baseline 20-pulsar array (Dataset A). The blue histogram corresponds to the null distribution ($H_0$, $\rho_{\mathrm{tot}}=0$). As the injected signal strength increases, the distribution of $\mathcal{T}$ shifts systematically towards higher values.

\begin{figure}[htbp]
\centering
\includegraphics[width=0.85\columnwidth]{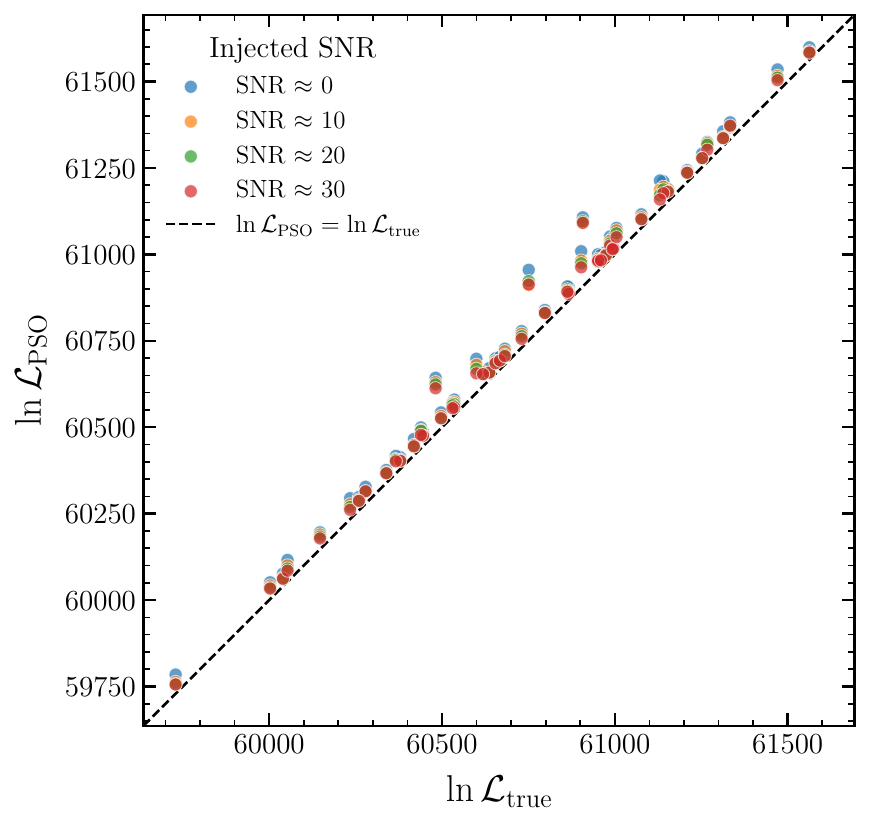}
\caption{
Comparison between the maximized log-likelihood identified with PSO ($\ln \mathcal{L}_{\mathrm{PSO}}$) and the likelihood at the true parameters ($\ln \mathcal{L}_{\mathrm{true}}$).
The dashed line indicates equality ($\ln \mathcal{L}_{\mathrm{PSO}} = \ln \mathcal{L}_{\mathrm{true}}$).}
\label{fig:likelihood_comparison}
\end{figure}

Although the systematic shift is evident, the null distribution exhibits a significant tail extending to high values. This behavior aligns with our theoretical expectations. Because our joint search navigates a vast parameter space that is undefined under the null hypothesis (e.g., frequency, sky location, and chirp mass), the look-elsewhere effect distorts the standard $\chi^2$ distribution into an EVD. Occasional quasi-sinusoidal structures in the unmodeled stochastic red noise can accidentally align across a subset of pulsars, leading the highly flexible CGW model to artificially inflate the likelihood. We verified that for all realizations, $\ln \mathcal{L}_{\mathrm{PSO}} \ge \ln \mathcal{L}_{\mathrm{true}}$ holds, confirming that the algorithm successfully converged to the global maximum, and the outliers represent genuine physical noise fluctuations rather than optimization failures (see Figure~\ref{fig:likelihood_comparison}).

\begin{figure}[htbp]
\centering
\includegraphics[width=0.98\columnwidth]{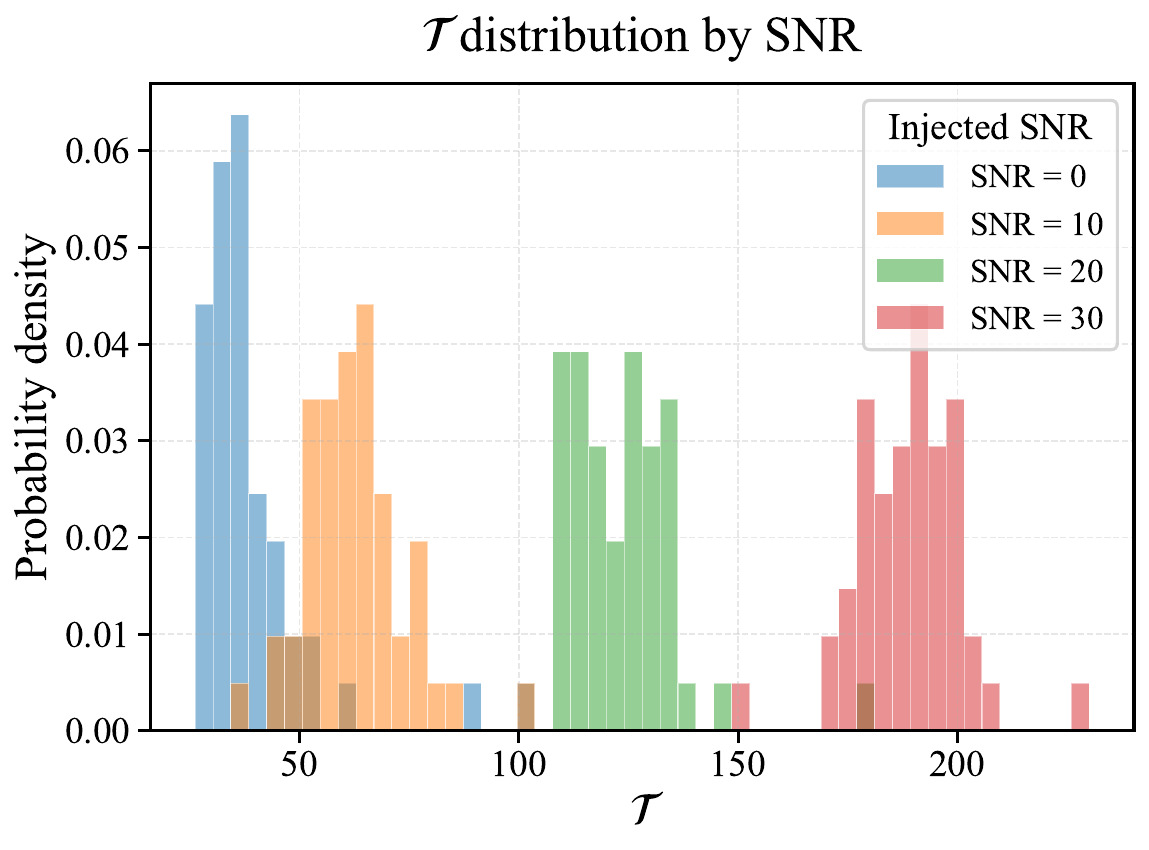}
\caption{
Distributions of the detection statistic $\mathcal{T}$ for four different signal strengths: SNR $\approx$ 0 (blue), 10 (orange), 20 (green), and 30 (red).
Each histogram is constructed from 50 independent realizations using the 30-pulsar array (Dataset B).}
\label{fig:snr_distribution_30psrs}
\end{figure}

Crucially, this noise-driven EVD tail is dramatically suppressed by increasing the array size. When analyzing the extended 30-pulsar array (Dataset B) under identical network SNRs, the high-value tail of the null distribution shrinks significantly, as shown in Figure~\ref{fig:snr_distribution_30psrs}.

This behavior reflects the increased number of spatial constraints supplied by the larger array. A genuine CGW requires a highly specific quadrupolar spatial correlation pattern across the sky. With 30 geographically distributed pulsars, the array becomes too heavily constrained for independent, uncorrelated stochastic red noise processes to accidentally mimic a coherent GW signal. Consequently, the empirical detection threshold (corresponding to an effective false alarm probability of $\alpha \approx 0.06$) dropped dramatically from $\mathcal{T}_{\mathrm{th,20}} = 127.0_{-29.1}^{+64.7}$ (68\% confidence interval $[97.9, 191.7]$) for the 20-pulsar array to $\mathcal{T}_{\mathrm{th,30}} = 58.0_{-3.9}^{+44.8}$ (68\% CI $[54.1, 102.8]$) for the 30-pulsar array. Here, the $1\sigma$ asymmetric uncertainties are rigorously determined using the exact binomial distribution for the empirical quantiles over the 50 noise realizations. This substantial reduction in both the threshold value and its statistical upper bound further emphasizes that expanding the array not only lowers the detection barrier but also strongly stabilizes the extreme-value tail of the background noise distribution.

As summarized in Table~\ref{tab:detection_rates}, this threshold reduction provides a significant improvement in detectability. While the 20-pulsar array only achieved a 10\% detection rate at $\rho_{\mathrm{tot}} \approx 20$, the 30-pulsar array reached a robust 100\% detection rate at the exact same signal strength. This empirically validates that expanding the number of monitored pulsars in next-generation observatories will naturally eradicate the false alarm penalty induced by the look-elsewhere effect in blind searches.
\begin{table}[htbp]
\centering
\caption{
Empirical detection fractions for different injected SNRs based on 50 realizations for the 20-pulsar and 30-pulsar datasets.
The thresholds correspond to an effective false-alarm probability of $0.06$.
}
\label{tab:detection_rates}
\renewcommand{\arraystretch}{1.4}
\begin{tabular}{ccc}
\hline
\hline
\textbf{Injected SNR} ($\rho_{\mathrm{tot}}$) & \textbf{$P_d$ (20 Pulsars)} & \textbf{$P_d$ (30 Pulsars)} \\
\hline
0 (Noise only) & $0.06$ (False Alarm) & $0.06$ (False Alarm) \\
10 & $0.06$ & $0.66$ \\
20 & $0.10$ & $1.00$ \\
30 & $0.90$ & $1.00$ \\
\hline
\hline
\end{tabular}
\end{table}

\subsubsection{CGW Parameter Recovery}
\begin{figure*}[t] 
    \centering
    \begin{minipage}{0.95\textwidth}
        \textbf{(a) Dataset A (20 Pulsars)} \par
        \vspace{0.1cm}
    \includegraphics[width=\linewidth]{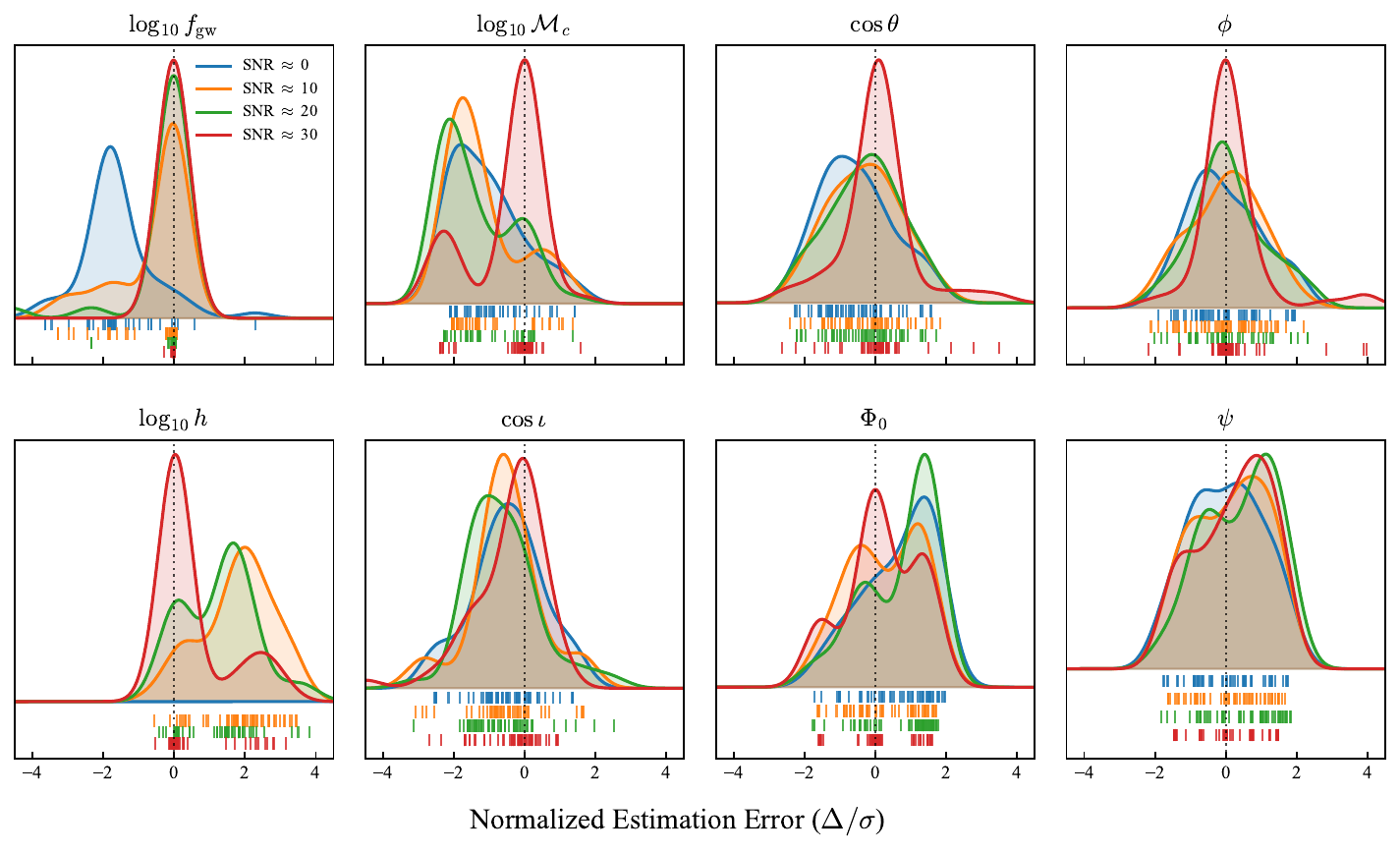}
    \end{minipage}
    \vspace{0.5cm}
    \begin{minipage}{0.95\textwidth}
        \textbf{(b) Dataset B (30 Pulsars)} \par
        \vspace{0.1cm}
\includegraphics[width=\linewidth]{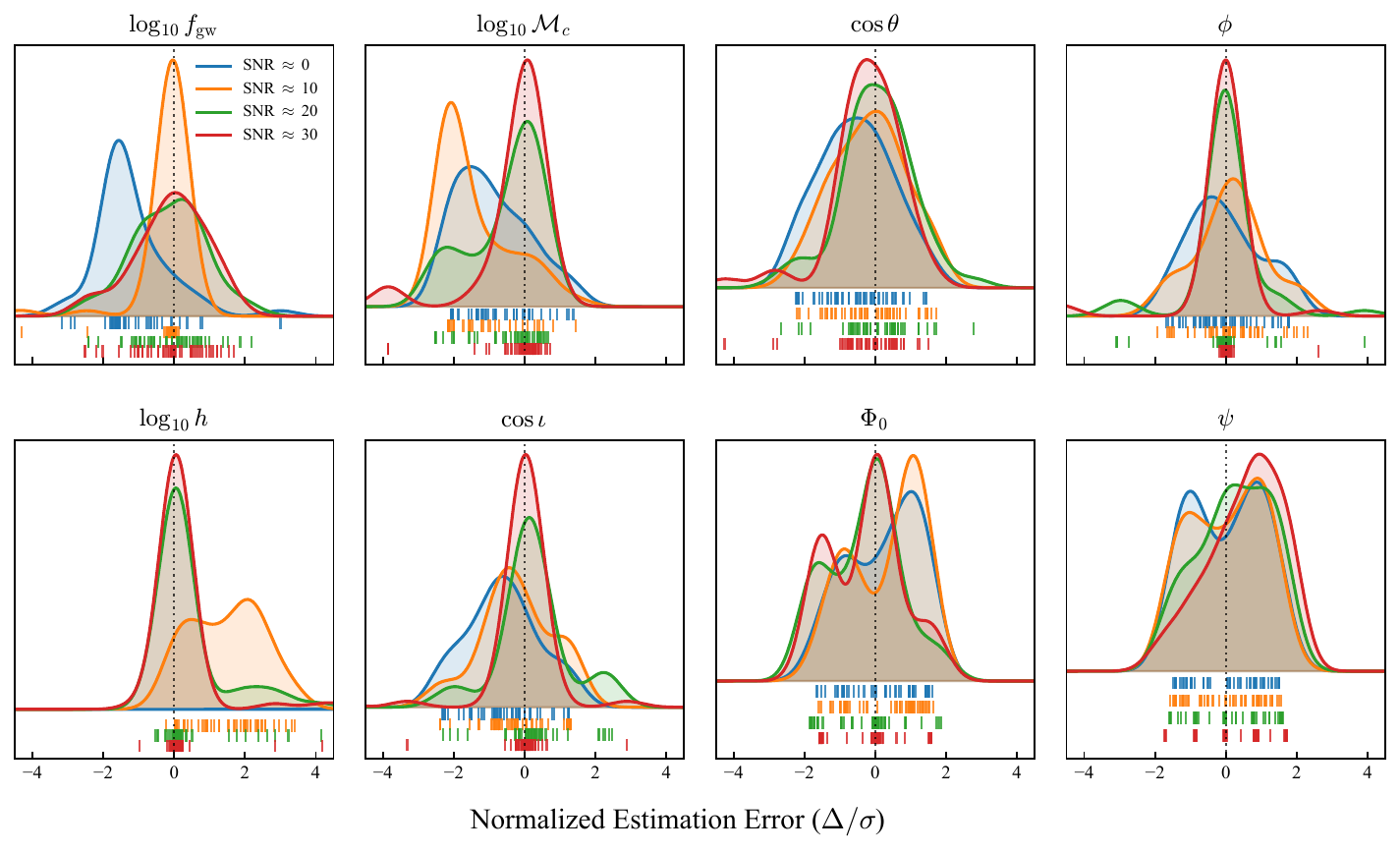} 
    \end{minipage}
    \caption{
    Comparison of CGW parameter estimation performance. 
    Top panel \textbf{(a)} displays the distributions for the 20-pulsar baseline array, and bottom panel \textbf{(b)} shows the extended 30-pulsar array. 
    The x-axes show the normalized estimation errors ($\Delta \lambda / \sigma_{\Delta\lambda}$), with solid lines representing the Gaussian kernel density estimates (KDE) and bottom rug plots marking individual realizations. 
    The transition from 20 to 30 pulsars effectively breaks the parameter degeneracies at moderate signal strengths (SNR $\approx$ 20, green), forcing the broad error envelopes of $\mathcal{M}_c$, $\log_{10} h$, and sky locations to converge into zero-centered distributions.
    }
    \label{fig:cgw_errors_comparison}
\end{figure*}

We define the estimation error for a parameter $\lambda$ as $\Delta \lambda = \lambda_{\mathrm{best}} - \lambda_{\mathrm{true}}$. To visualize the estimation accuracy while accounting for scale variations across different signal strengths, Figure~\ref{fig:cgw_errors_comparison} displays the kernel density estimates (KDE) of the normalized estimation errors ($\Delta \lambda / \sigma_{\Delta\lambda}$) for both the 20-pulsar baseline (panel a) and the 30-pulsar array (panel b). 
Rug plots are included along the $x$-axis to indicate the underlying individual realizations. The comprehensive statistical results are summarized in Table~\ref{tab:est_stats_comparison_6groups}.
Several key characteristics of the search algorithm which emerge from this normalized representation are outlined below.

First, for the gravitational-wave frequency ($\log_{10} f_{\mathrm{gw}}$), the SNR $\approx 10$ distribution in the 30-pulsar array appears exceptionally narrow near zero. This should not be interpreted as uniformly precise low-SNR frequency recovery. At this transitional SNR, realizations that pass the detection threshold lock onto the injected signal and cluster near the true frequency, while those that remain below threshold can be dominated by noise-induced likelihood peaks and therefore yield biased frequency estimates. Because the errors are normalized by the empirical width $\sigma_{\Delta\lambda}$, these biased realizations enlarge the normalization scale and visually compress the successfully recovered cases into a narrow central peak.

Second, angular parameters such as the initial phase $\Phi_0$ and polarization angle $\psi$ exhibit multimodal distributions even at high signal strengths (SNR $\approx 30$, red). These secondary peaks reflect the expected discrete phase--polarization degeneracies of the CGW likelihood surface, confirming that the PSO algorithm successfully identifies the equivalent degenerate states without artificial confinement.

The comparative analysis between panels (a) and (b) directly correlates with the detection probabilities reported in Table~\ref{tab:detection_rates}. At moderate signal strength ($\rho_{\mathrm{tot}} \approx 20$), the 20-pulsar array has only a 10\% detection probability, and the corresponding distributions for parameters such as the chirp mass ($\mathcal{M}_c$), strain amplitude ($h$), and sky location ($\cos\theta, \phi$) are visibly shifted away from zero rather than centered on the injected values. This indicates that most realizations do not lock onto the true CGW signal; instead, the likelihood maximum is often set by a noise-driven or incorrectly localized peak. Expanding the array to 30 pulsars provides the critical spatial baselines required to break these geometric and $h$-$\mathcal{M}_c$ degeneracies. Consequently, at the same SNR $\approx 20$, the normalized error distributions become much more concentrated around zero, consistent with the 100\% detection rate and enabling more robust 3D source localization.

\begin{table*}[htbp]
\centering
\caption{
Comparison of parameter estimation accuracy for Dataset A (20 pulsars) and Dataset B (30 pulsars) across different signal strengths (SNR $\approx$ 10, 20, 30).
Values represent the median bias and $1\sigma$ uncertainty (estimated as half of the central 68\% interval) derived from 50 independent realizations.
}
\label{tab:est_stats_comparison_6groups}
\renewcommand{\arraystretch}{1.2}
\setlength{\tabcolsep}{3pt}
\resizebox{\textwidth}{!}{%
\begin{tabular}{l|cc|cc|cc|cc|cc|cc}
\hline\hline
\multirow{3}{*}{Parameter} & \multicolumn{6}{c|}{Dataset A (20 Pulsars)} & \multicolumn{6}{c}{Dataset B (30 Pulsars)} \\
\cline{2-13}
 & \multicolumn{2}{c|}{SNR $\approx$ 10} & \multicolumn{2}{c|}{SNR $\approx$ 20} & \multicolumn{2}{c|}{SNR $\approx$ 30} & \multicolumn{2}{c|}{SNR $\approx$ 10} & \multicolumn{2}{c|}{SNR $\approx$ 20} & \multicolumn{2}{c}{SNR $\approx$ 30} \\
 & Bias & Unc. & Bias & Unc. & Bias & Unc. & Bias & Unc. & Bias & Unc. & Bias & Unc. \\
\hline
$\log_{10} f_{\mathrm{gw}}$ & $-8.6\times 10^{-3}$ & $2.8\times 10^{-1}$ & $-2.3\times 10^{-3}$ & $6.3\times 10^{-3}$ & $-5.8\times 10^{-4}$ & $3.4\times 10^{-3}$ & $-3.8\times 10^{-3}$ & $1.5\times 10^{-2}$ & $-3.2\times 10^{-4}$ & $4.9\times 10^{-3}$ & $-1.8\times 10^{-4}$ & $3.6\times 10^{-3}$ \\
$\log_{10} \mathcal{M}_c$ & $-0.84$ & $0.67$ & $-0.87$ & $0.59$ & $-0.03$ & $0.59$ & $-1.13$ & $0.54$ & $-0.05$ & $0.56$ & $0.01$ & $0.12$ \\
$\log_{10} D_L$ & $-2.78$ & $1.52$ & $-2.72$ & $1.73$ & $-0.08$ & $1.68$ & $-2.68$ & $1.72$ & $-0.09$ & $1.18$ & $-0.01$ & $0.23$ \\
$\cos\theta$ & $-0.13$ & $0.48$ & $-0.12$ & $0.51$ & $0.02$ & $0.09$ & $-0.11$ & $0.51$ & $0.01$ & $0.16$ & $-0.02$ & $0.09$ \\
$\phi$ (rad) & $0.04$ & $1.55$ & $-0.07$ & $1.34$ & $-0.01$ & $0.11$ & $0.20$ & $1.18$ & $0.01$ & $0.10$ & $-0.01$ & $0.04$ \\
$\log_{10} h$ & $1.23$ & $0.77$ & $1.18$ & $0.78$ & $0.05$ & $0.62$ & $1.13$ & $0.77$ & $0.05$ & $0.18$ & $0.01$ & $0.04$ \\
$\cos\iota$ & $-0.25$ & $0.24$ & $-0.20$ & $0.22$ & $-0.02$ & $0.13$ & $-0.15$ & $0.57$ & $0.05$ & $0.15$ & $0.01$ & $0.05$ \\
$\Phi_0$ (rad) & $0.36$ & $1.97$ & $2.07$ & $1.75$ & $0.09$ & $2.05$ & $1.39$ & $2.10$ & $-0.03$ & $1.61$ & $0.06$ & $1.79$ \\
$\psi$ (rad) & $0.47$ & $2.17$ & $1.00$ & $1.85$ & $1.14$ & $2.75$ & $0.30$ & $2.31$ & $0.43$ & $2.32$ & $1.51$ & $2.29$ \\
\hline\hline
\end{tabular}%
}
\end{table*}

\begin{figure}[htbp]
\centering
\includegraphics[width=1.0\columnwidth]{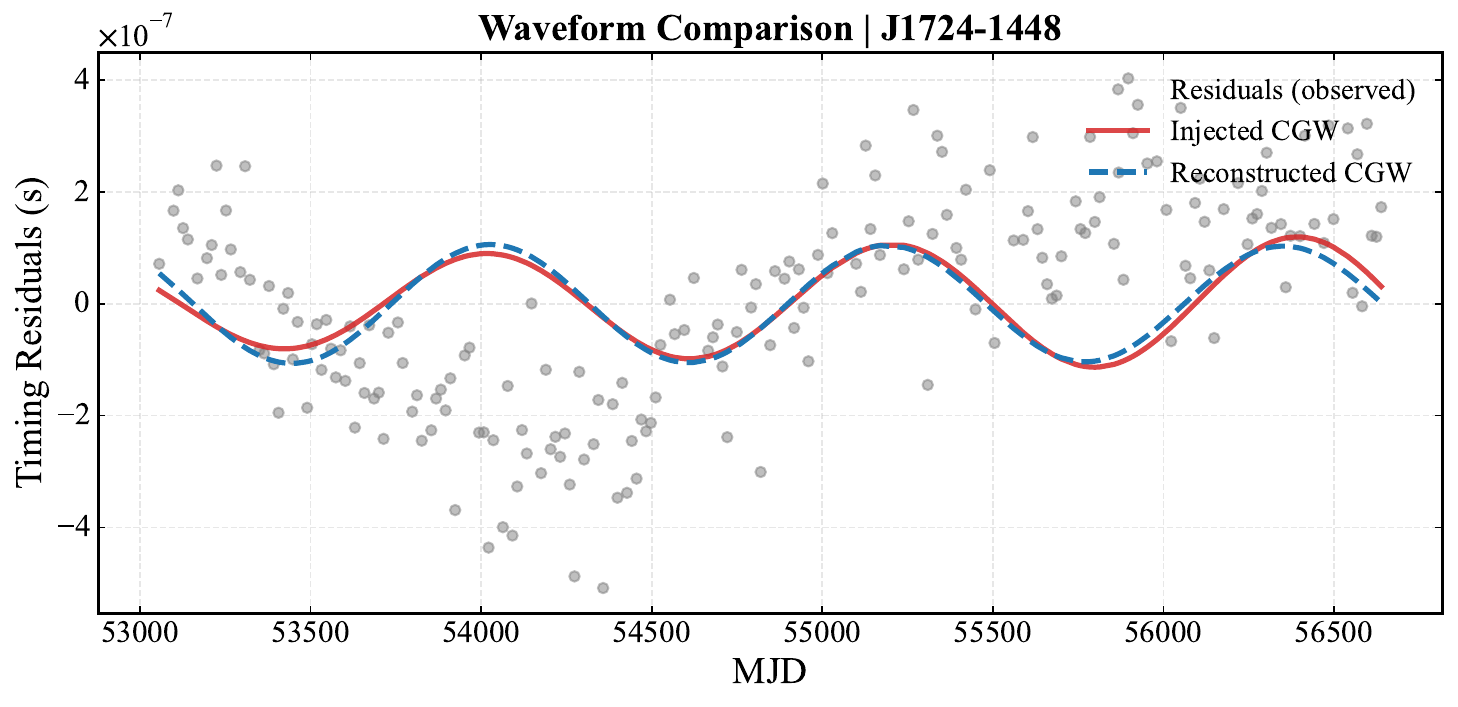}
\caption{
Time-domain reconstruction of the isolated CGW signal for pulsar J1724$-$1448 from a representative 20-pulsar array realization at $\rho_{\mathrm{tot}}\approx30$.
Gray points denote the observed timing residuals including noise, the solid red curve is the injected CGW waveform, and the dashed blue curve is the maximum-likelihood reconstructed waveform.
}
\label{fig:cgw_waveform_example}
\end{figure}

As a direct time-domain check of the signal reconstruction, Figure~\ref{fig:cgw_waveform_example} shows the isolated CGW waveform for one representative pulsar, J1724$-$1448, in the 20-pulsar array at $\rho_{\mathrm{tot}}\approx30$. The recovered waveform follows the injected CGW phase and amplitude closely despite the much larger stochastic timing residuals, illustrating that the joint fit can extract the deterministic signal component while simultaneously accounting for pulsar red noise.

\subsubsection{Red Noise Characterization}
\label{subsubsec:red_noise_characterization}

A central motivation for the joint-search framework is to characterize the covariance between a deterministic continuous gravitational wave (CGW) signal and intrinsic pulsar red noise within the same likelihood analysis. Even when red-noise and CGW parameters are sampled simultaneously in a Bayesian stage, a subsequent frequentist statistic may still be evaluated using a covariance matrix that is not separately profiled under the signal and noise-only hypotheses. To diagnose the impact of this distinction, we examine the recovered intrinsic red-noise parameters under the full joint $H_1$ optimization and compare them with a noise-only $H_0$ optimization in which the deterministic CGW template is omitted.

\begin{figure}[htbp]
\centering
\includegraphics[width=0.9\columnwidth]{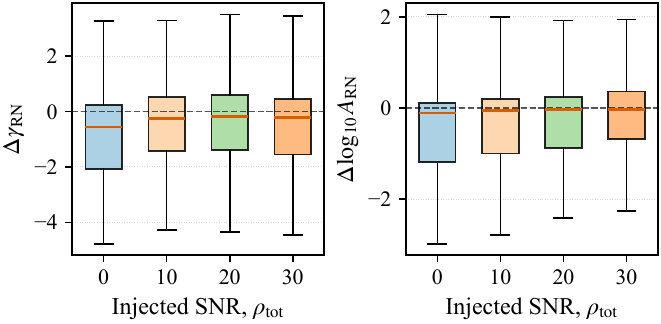}
\caption{
$H_1$-only red-noise parameter recovery as a function of injected array SNR.
The boxplots show the aggregate errors in the recovered spectral index
$\Delta\gamma_{\rm RN}$ and amplitude $\Delta\log_{10}A_{\rm RN}$ over all pulsars and noise realizations.
The weak decrease in scatter with increasing SNR reflects improved separation between the coherent CGW signal and the stochastic intrinsic red-noise components.
}
\label{fig:rn_boxplot}
\end{figure}

We first consider the red-noise parameters recovered by the joint $H_1$ fit. For each pulsar, we compare the recovered spectral index $\gamma_{\rm RN}$ and amplitude $\log_{10} A_{\rm RN}$ with their injected values. Figure~\ref{fig:rn_boxplot} summarizes the aggregate estimation errors over all pulsars and all noise realizations as a function of injected array SNR. The overall scatter is broad, as expected, because many pulsars have intrinsically weak red noise and their individual red-noise hyperparameters are only weakly constrained by a finite data realization. Nevertheless, the $H_1$-only errors show a mild reduction with increasing signal strength. This behavior is physically expected: as the injected CGW becomes stronger, the coherent deterministic signal is more accurately localized and reconstructed by the joint model, reducing the tendency of the CGW template to absorb stochastic low-frequency red-noise fluctuations. The remaining residuals therefore provide a cleaner estimate of the intrinsic red-noise covariance.

\begin{figure}[htbp]
\centering
\includegraphics[width=0.9\columnwidth]{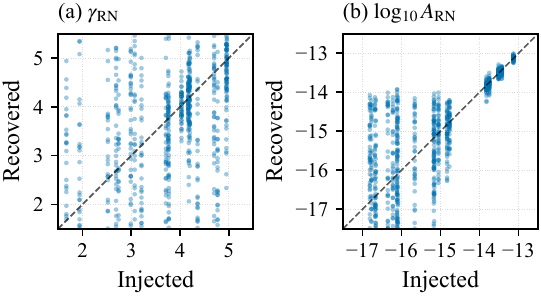}
\caption{
Recovered versus injected intrinsic red-noise parameters for individual pulsars in the 20-pulsar array at $\rho_{\rm tot}\approx30$.
Left: spectral index $\gamma_{\rm RN}$.
Right: amplitude $\log_{10} A_{\rm RN}$.
The dashed diagonal indicates perfect recovery.
Pulsars with stronger injected red noise are more tightly constrained, whereas weak-red-noise pulsars show larger realization-to-realization scatter.
}
\label{fig:rn_scatter}
\end{figure}

At the level of individual pulsars, the recovered parameters also track the injected red-noise population. Figure~\ref{fig:rn_scatter} compares the recovered and injected red-noise parameters for the 20-pulsar array at $\rho_{\rm tot}\approx30$. Pulsars with strong injected red noise are recovered most accurately, while pulsars with very weak red noise exhibit larger scatter. This behavior is expected: when the intrinsic red-noise amplitude is small, the likelihood is relatively flat in the red-noise hyperparameters, and different stochastic realizations can support a broad range of $(A_{\rm RN},\gamma_{\rm RN})$ values without significantly changing the total likelihood.

To more directly diagnose signal--noise leakage, we compare the full joint $H_1$ fit against a noise-only $H_0$ fit. Rather than examining $\gamma_{\rm RN}$ and $A_{\rm RN}$ separately, we evaluate the induced error in the red-noise power spectral density at the injected CGW frequency,
\begin{equation}
\Delta \log_{10} S_{\rm RN}(f_{\rm gw})
=
\log_{10} S_{\rm RN}^{\rm rec}(f_{\rm gw})
-
\log_{10} S_{\rm RN}^{\rm inj}(f_{\rm gw}) .
\end{equation}
For the power-law red-noise model in Eq.~\eqref{eq:rn_psd}, constants cancel in this difference, yielding
\begin{equation}
\Delta \log_{10} S_{\rm RN}(f_{\rm gw})
=
2\Delta \log_{10} A_{\rm RN}
-
\Delta\gamma_{\rm RN}
\log_{10}\left(\frac{f_{\rm gw}}{f_{\rm yr}}\right) .
\label{eq:rn_power_error}
\end{equation}
This quantity directly measures whether the red-noise covariance is inflated or suppressed at the frequency where the deterministic CGW signal contributes power.

\begin{figure*}[t]
\centering
\includegraphics[width=0.95\textwidth]{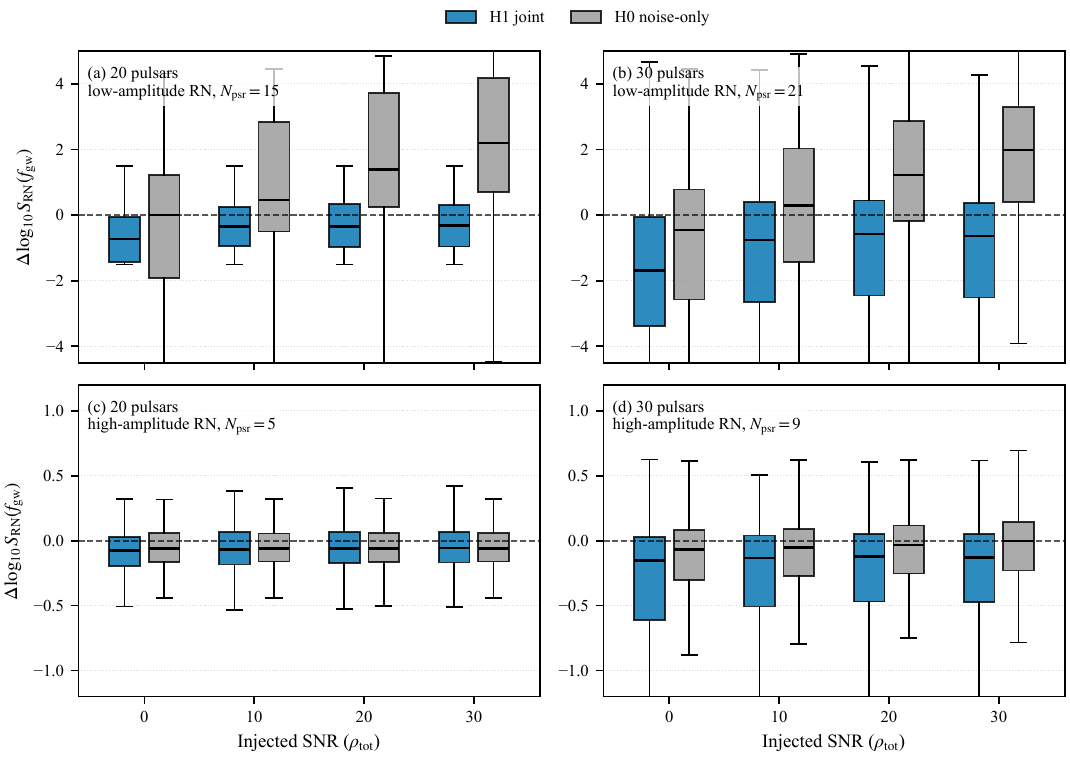}
\caption{
Signal--noise leakage diagnostic for red-noise characterization.
The plotted quantity is the recovered red-noise power error at the injected CGW frequency, $\Delta\log_{10}S_{\rm RN}(f_{\rm gw})$.
Results are shown for the 20-pulsar and 30-pulsar datasets, with pulsars divided into low-amplitude and high-amplitude red-noise subsets using $\log_{10}A_{\rm RN}=-14.5$.
Blue boxes denote the full joint $H_1$ fit, and gray boxes denote the noise-only $H_0$ fit.
}
\label{fig:rn_leakage}
\end{figure*}

Figure~\ref{fig:rn_leakage} shows the resulting leakage diagnostic for both the 20-pulsar and 30-pulsar datasets. Pulsars are separated into low-amplitude and high-amplitude red-noise subsets according to their injected amplitudes, using $\log_{10}A_{\rm RN}=-14.5$ as the dividing value. The noise-only $H_0$ fit exhibits the expected signal-to-noise leakage: for low-amplitude red-noise pulsars, the recovered red-noise power develops a positive bias that grows with the injected CGW SNR. This demonstrates that, when the deterministic CGW template is omitted, part of the CGW power can be absorbed by the red-noise model, artificially inflating the inferred red-noise covariance.

By contrast, the joint $H_1$ estimates do not show the same SNR-dependent positive drift. This indicates that profiling the CGW signal and red-noise covariance together under $H_1$ suppresses the specific bias seen in the $H_0$ fit, namely the leakage of deterministic CGW power into the red-noise model. The $H_1$ estimates are not unbiased in every realization: their distributions are mildly shifted toward negative values and can develop a negative tail, especially for pulsars whose intrinsic red noise is weak. This reflects the opposite leakage channel. In a finite noise realization, stochastic low-frequency red-noise fluctuations can occasionally be fitted by the flexible deterministic CGW template, leaving too little power to be assigned to the intrinsic red-noise covariance. As the signal SNR increases, the coherent CGW component is better determined, and this accidental noise-to-signal leakage is reduced, consistent with the $H_1$-only recovery trends in Figure~\ref{fig:rn_boxplot}.

For high-amplitude red-noise pulsars, both the $H_0$ and $H_1$ estimates remain close to zero. In these pulsars the intrinsic red-noise process is sufficiently strong to be directly constrained by the timing residuals, so neither the omission nor the inclusion of the deterministic CGW template substantially biases the recovered red-noise power. We use these boxplots as a qualitative diagnostic of the dominant leakage direction within each simulated dataset, rather than as a controlled comparison of red-noise parameter precision between the 20-pulsar and 30-pulsar configurations. The relevant comparison is therefore not the width of the distributions across the two datasets, but the contrast between $H_0$ and $H_1$ within each dataset: $H_0$ shows an SNR-dependent positive bias from signal-to-noise leakage, whereas $H_1$ largely suppresses this positive drift but can retain a smaller negative bias or tail from noise-to-signal leakage in weak-red-noise pulsars.

Overall, these results demonstrate that the joint estimation framework does not merely recover the deterministic CGW signal, but also suppresses the red-noise covariance from being systematically inflated by unmodeled signal power. The remaining scatter in the $H_1$ red-noise estimates reflects finite realization noise and the weak identifiability of subdominant intrinsic red-noise components, rather than an SNR-dependent leakage of CGW power into the noise model.

\subsection{Comparison with the $\mathcal{F}_p$ statistic}
\label{subsec:fp_comparison}

As a final diagnostic, we compare the $\mathcal{T}$ statistic with the standard pulsar-term $\mathcal{F}_p$ statistic for the 30-pulsar data set. The purpose is not to claim uniform superiority over an optimized $\mathcal{F}_p$ pipeline, but to show how strongly the practical performance of $\mathcal{F}_p$ depends on the covariance used in its inner product. For this diagnostic we use an enlarged 100-realization ensemble, which differs from the preceding 50-realization ensembles only by the additional noise realizations. This larger sample is used only to draw smoother receiver operating characteristic (ROC) curves. 
We first compare the detection performance of $\mathcal{T}$ and $\mathcal{F}_p$ using empirical ROC curves, and then assess the false-alarm probabilities obtained with analytical $\mathcal{F}_p$ thresholds. In both comparisons, we evaluate $\mathcal{F}_p$ using three covariance choices: the injected true covariance (as an oracle reference, not attainable in practice), the covariance obtained from the selected $H_0$ fit, and the covariance obtained from the selected $H_1$ fit.

\subsubsection{Empirical ROC curves}
\label{subsec:emproc}

ROC curves characterize the trade-off between detection probability and false-alarm probability and have been used to compare GW detection strategies~\cite{2003PhRvD..68j2001A,2011IJMPD..20.2051D}. In the present comparison, empirical ROC curves are particularly useful because estimating the noise covariance from the data can alter the null distribution of the detection statistic.
Determining the threshold for each statistic from its corresponding noise-only realizations therefore allows detection performance to be compared at the same empirically measured false-alarm probability.

\begin{figure}[t]
\centering
\includegraphics[width=\columnwidth]{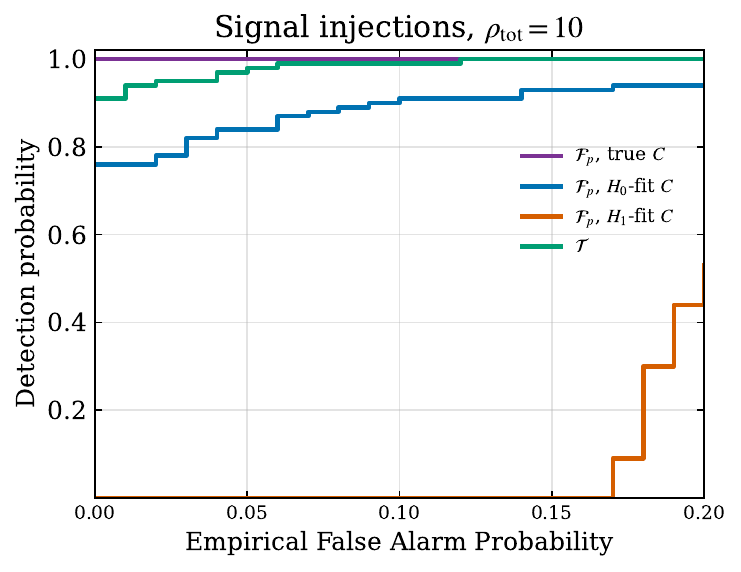}
\caption{Empirically determined ROC curves for the 30-pulsar simulations with $\rho_{\rm tot}=10$. For each curve, the threshold is obtained from the corresponding 100 noise-only realizations, and the detection probability is obtained from the 100 signal-injected realizations. The horizontal coordinate is therefore the empirically measured false-alarm probability rather than the nominal value derived from an analytic null distribution. The true covariance $\mathcal{F}_p$ curve is included as an oracle reference.}
\label{fig:fp_roc_30psr_empirical}
\end{figure}

In Fig.~\ref{fig:fp_roc_30psr_empirical}, we show empirical ROC curves for the 30-pulsar ensemble with $\rho_{\rm tot}=10$, where the covariance effect is most visible. 
For each statistic, the false-alarm probability is estimated as the fraction of the corresponding 100 noise-only realizations exceeding a given threshold, and the detection probability as the fraction of the 100 signal-injected realizations exceeding the same threshold. For each fitted-covariance $\mathcal{F}_p$ curve, the same covariance estimation procedure is applied to both ensembles. Varying the threshold yields the ROC curve; thus, both coordinates are determined empirically. With 100 noise-only realizations, the empirical false-alarm probability has a resolution of $0.01$.

At an empirical false-alarm probability of $0.06$, the $\mathcal{T}$ statistic detects $99/100$ realizations, compared with $100/100$, $87/100$, and $0/100$ for the $\mathcal{F}_p$ statistics evaluated using the true, $H_0$-fit, and $H_1$-fit covariances, respectively.
The low detection probability of the $H_1$-fit $\mathcal{F}_p$ statistic in this ensemble results from the pronounced high statistic tail of its noise-only distribution.
Consequently, setting its empirical false-alarm probability to $0.06$ requires a threshold larger than all of its values in the $\rho_{\rm tot}=10$ ensemble.

\subsubsection{Assessing analytical $\mathcal{F}_p$ thresholds}
\label{subsec:emproc}

\begin{figure*}[t]
\centering
\includegraphics[width=0.98\textwidth]{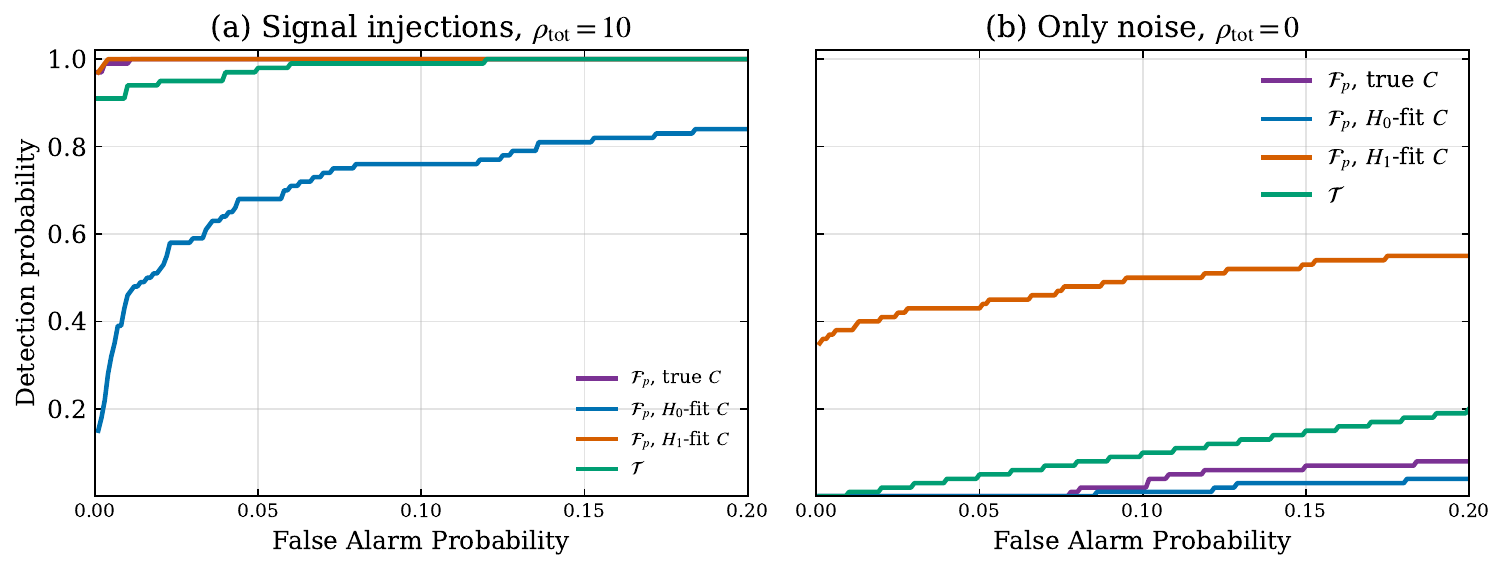}
\caption{
ROC-curve comparison between $\mathcal{T}$ and $\mathcal{F}_p$ for the 30-pulsar simulations.
The vertical axis gives the detection probability in the signal-injection panel and the false-detection fraction in the only-noise panel.
Left: weakest nonzero injections, $\rho_{\rm tot}=10$.
Right: only-noise realizations, $\rho_{\rm tot}=0$.
The $\mathcal{F}_p$ curves use the analytic fixed-covariance threshold with a 31-bin trials correction, and the $\mathcal{T}$ curve uses the noise-only $\mathcal{T}$ threshold.
The true-covariance $\mathcal{F}_p$ curve is an oracle reference.
}
\label{fig:fp_roc_30psr}
\end{figure*}

The empirical ROC curves above compare detection performance at the same measured false-alarm probability. We now assess whether the standard analytical $\mathcal{F}_p$ thresholds yield the nominal false-alarm probabilities when the covariance is estimated from the data. Figure~\ref{fig:fp_roc_30psr} shows detection and false-detection fractions as functions of the nominal false-alarm probability, using the same signal-injected and noise-only ensembles.

Following Refs.~\cite{Ellis_2012,sardesai_optimal_2026}, we evaluate $\mathcal{F}_p$ on an independent Fourier grid with spacing $1/T_{\rm obs}$ over the same frequency range used in the main search. This gives 31 frequency bins. For a target total false-alarm probability $\alpha$, the $\mathcal{F}_p$ threshold is obtained from the analytic fixed-covariance null distribution with the 31-bin trials correction,
$1-[F_{\chi^2_{60}}(2\mathcal{F}_{p,{\rm th}})]^{31}=\alpha$.
At $\alpha=0.06$, this gives $\mathcal{F}_{p,{\rm th}}=48.209$.

The curves in Fig.~\ref{fig:fp_roc_30psr} differ from a conventional empirical ROC curve shown in Fig.~\ref{fig:fp_roc_30psr_empirical}. The horizontal coordinate is not the measured false-positive fraction obtained from a common score threshold. Instead, it is the target false-alarm probability used to set the threshold for each statistic. The $\mathcal{F}_p$ curves use the analytic fixed-covariance threshold at each target $\alpha$, whereas the $\mathcal{T}$ curve uses the threshold $\mathcal{T}_{\rm th}(\alpha)$ obtained from the noise-only $\mathcal{T}$ distribution described in Sec.~\ref{subsec:threshold_method}. Thus, the left panel gives the detection probability for weak-signal injections, while the right panel gives the false-detection fraction for zero-signal data.
False-detection occurs when the calculated $\mathcal{F}_p$ exceeds the  $\mathcal{F}_{p,{\rm th}}$ derived from the theoretical false alarm probability.

The left panel shows the weak-signal cost of using a noise-only covariance in $\mathcal{F}_p$. With the true injected covariance, $\mathcal{F}_p$ detects $100/100$ realizations at $\alpha=0.06$ and $99/100$ realizations at $\alpha=0.005$, confirming that the weak injections are visible to the statistic when the covariance is known. With the $H_0$-fit covariance, the corresponding detection counts drop to $71/100$ at $\alpha=0.06$ and $32/100$ at $\alpha=0.005$. The $\mathcal{T}$ statistic detects $99/100$ and $91/100$ realizations at the same two target false-alarm probabilities. This behavior is consistent with the leakage diagnostic above: when the deterministic CGW template is omitted in the $H_0$ fit, part of the weak signal can be absorbed into the red-noise covariance before $\mathcal{F}_p$ is evaluated. 

The $H_1$-fit covariance behaves differently. For signal-containing data it restores the lost weak-signal power, giving $100/100$ detections at $\alpha=0.06$ for $\rho_{\rm tot}=10$. However, the right panel shows that this improvement cannot be interpreted using the usual fixed-covariance $\mathcal{F}_p$ threshold. In zero-signal data, the $H_1$-fit $\mathcal{F}_p$ statistic gives $45/100$ false detections at the same target $\alpha=0.06$, and already gives $37/100$ false detections at $\alpha=0.005$. This occurs because the covariance has been inferred under the signal hypothesis from the same data: some red-noise realizations can be fitted in a signal-like way, and the resulting covariance can increase $\mathcal{F}_p$ even when no CGW signal is present. This is the same noise-to-signal leakage identified in Fig.~\ref{fig:rn_leakage}. In a noise-only realization, the flexible CGW template under $H_1$ can absorb a low-frequency red-noise fluctuation, leaving less power to be attributed to intrinsic red noise than under $H_0$. Consequently, the $H_1$-fit covariance underestimates the low-frequency noise power. When this covariance is used to compute $\mathcal{F}_p$, the inverse-covariance weighting makes the remaining low-frequency residuals appear more significant and thereby inflates $\mathcal{F}_p$. This explains both the difference between the maximum likelihood red noise parameters obtained under $H_1$ and $H_0$ and the elevated false-detection fraction of the $H_1$-fit covariance curve. 
Because the covariance and CW template are inferred from the same realization, this curve is diagnostic, and the standard analytic $\mathcal{F}_p$ threshold does not yield the nominal false-alarm probability for this construction.

We examined the maximum-likelihood CW parameters for the 37 false detections at $\alpha=0.005$, for which $\mathcal{F}_{p,{\rm th}}=53.759$. Of these, 35 have their maximum $\mathcal{F}_p$ in the lowest Fourier bin, and in 35 cases that peak lies within $0.06$ dex of the CW frequency selected by the $H_1$ fit. The medians (16th--84th percentiles) are $\log_{10}(f_{\rm gw}/{\rm Hz})=-8.493$ ($-8.503$ to $-8.469$), $\log_{10}h=-12.89$ ($-13.71$ to $-12.24$), and $\log_{10}(\mathcal{M}/M_\odot)=8.39$ ($8.00$ to $8.99$). The recovered frequencies are therefore concentrated near the lowest characteristic frequency, $1/T_{\rm obs}$, whereas the fitted angular parameters (sky position and orientation) are broadly distributed across their search ranges. Thus these fits select noise-dependent, spurious low-frequency templates rather than a common physical source.

By contrast, the true-covariance and $H_0$-fit $\mathcal{F}_p$ curves remain low in the zero-signal panel, giving no false detections at $\alpha=0.06$ in this finite ensemble. The $\mathcal{T}$ statistic is defined as a comparison between the separately maximized $H_1$ and $H_0$ likelihoods, and its threshold is read from the corresponding noise-only distribution. In the zero-signal ensemble, this gives the expected operating behavior: $6/100$ false detections at $\alpha=0.06$. Thus the $H_0$-fit $\mathcal{F}_p$ result is more conservative in this ensemble, but this conservatism is accompanied by the weaker signal efficiency seen in the left panel.

The main lesson of Fig.~\ref{fig:fp_roc_30psr} is therefore not that $\mathcal{T}$ universally dominates $\mathcal{F}_p$, but that covariance estimation and threshold interpretation are inseparable. A noise-only covariance can preserve the analytic fixed-covariance interpretation but reduce weak-signal sensitivity. A signal-conditioned covariance can recover signal power but can also introduce signal-like red-noise contributions, so its false-detection fraction no longer represents the target false-alarm probability specified by the standard fixed-covariance formula. The $\mathcal{T}$ statistic avoids choosing a single covariance for both hypotheses by comparing the profiled likelihoods under $H_1$ and $H_0$ directly.

\subsection{Computational Efficiency}

In a numerical search that explicitly includes the complete pulsar term, the CGW parameter space contains $8+2N_p$ parameters: eight global CGW parameters, $N_p$ pulsar distances, and $N_p$ pulsar phases. Analytical maximization over the phases reduces this dimension to $8+N_p$. Both dimensions remain linear in $N_p$. The computational benefit is therefore the removal of $N_p$ highly multimodal phase directions from the numerical optimization, rather than a change in the asymptotic parameter-count scaling. Pulsar-specific noise parameters are included in both cases; their number depends on the adopted noise model, and they are unaffected by the phase maximization.

For the extended 30-pulsar array, one full PSO joint-estimation run with $40$ particles and $2000$ iterations completed in approximately $7.1$--$7.6 \times 10^2\,\mathrm{s}$ in our reference local-Docker benchmark. This benchmark was run inside a Docker container on an Apple M2 Mac. The container saw 8 CPU cores and the host had 16 GB of system memory. The implementation used Python 3.12 with the NumPy backend; CuPy/GPU acceleration was not used.

\section{Conclusions}\label{Sec:conc}
In this paper, we have presented a highly efficient, purely frequentist time-domain search algorithm for detecting CGW from individual supermassive black hole binaries using PTAs. By mathematically decoupling the pulsar distance from the highly uncertain pulsar phase, we transformed the phase maximization into a deterministic quartic polynomial root-finding problem. This analytical reduction allows us to dynamically eliminate the phase parameters from the numerical search space. Combined with PSO, our framework enables the simultaneous, joint estimation of the full multi-dimensional parameter space, encompassing both the global CGW parameters and the intrinsic red noise properties of individual pulsars.

Our simulation studies highlight three distinct advantages of this methodology relative to existing frequentist and Bayesian--frequentist continuous-wave search strategies: (1) by profiling the red-noise covariance separately under the signal and noise-only hypotheses, the method suppresses the systematic absorption of low-frequency CGW power into the red-noise model; (2) unlike standard incoherent statistics that often rely on monochromatic approximations, our time-domain algorithm explicitly integrates zeroth post-Newtonian radiation-reaction chirping, preserving phase coherence over the entire observation window; and (3) the analytical maximization over pulsar phases reduces the CGW search dimension from $8+2N_p$ to $8+N_p$ and removes $N_p$ highly multimodal phase directions from the numerical optimization.

Because the joint search explores a large non-linear parameter space, the resulting detection statistic does not have a simple analytical null distribution such as the fixed-noise $\chi^2$ distributions associated with standard $\mathcal{F}$-statistics. We therefore estimate the background distribution empirically using Monte Carlo noise replications. This is a common practice in GW data analysis when idealized theoretical threshold assumptions are not reliable for realistic data. The additional computational cost is the price of calibrating the statistic under the same joint-search procedure used for candidate events.

For an application to real PTA data, the noise parameters would first be fitted under $H_0$. An ensemble of noise-only data sets would then be generated from this fitted noise model using the actual observing setup, and the complete $H_1$ and $H_0$ searches would be repeated for every realization. The resulting distribution of $\mathcal{T}$ would determine the detection threshold. Thus the calibration requires an ensemble of data-matched noise-only simulations, rather than a single noise-only version of the observed data.

The current study is a methodological validation based on simulated datasets, and several physical complexities remain to be incorporated before deployment on real PTA data. These include chromatic noise processes such as dispersion-measure variations and scattering, common red processes across the array including a stochastic GWB, and waveform effects such as eccentricity and black hole spin beyond the circular-orbit model used here. Once these ingredients are included, the method can be applied to empirical PTA data sets. The linear scaling of the likelihood evaluation with the number of pulsars suggests that this framework may be useful for high-throughput searches in future PTA data sets containing many precisely timed pulsars, including those expected in the SKA era. This scalability advantage must be assessed together with the cost of empirical background estimation, since detection significances still require Monte Carlo noise replications.

\section*{Data Availability}
The data and analysis code that support the findings of this article are openly available \cite{Tao2026Zenodo}.

\acknowledgments
Y.W. gratefully acknowledges support from the National Natural Science Foundation of China (NSFC) under Grants No. 12673046, the National Key Research and Development Program of China (No. 2023YFC2206702 and No. 2022YFC2205201), and Major Science and Technology Program of Xinjiang Uygur Autonomous Region (No. 2022A03013-4). 
We acknowledge the High Performance Computing Platform at Huazhong University of Science and Technology for providing computational resources. The authors also acknowledge the Max Planck Institute for Gravitational Physics (Albert Einstein Institute) in Hannover for the use of the Atlas high-performance computing cluster. We are grateful to Bruce Allen and Rutger van Haasteren for helpful and insightful discussions.
Xuan Tao acknowledges the financial support from the China Scholarship Council (File No. 202506160011).
The authors thank the anonymous referee for helpful comments and suggestions.

\bibliography{PTAelove.bib}
\appendix
\section{Simulated Pulsar Parameters and Signal-to-Noise Ratios}
\label{app:pulsar_params}

This appendix details the noise properties and signal contributions for the simulated PTAs.
Table~\ref{tab:pulsar_combined} lists the injected red noise parameters ($\log_{10} A_{\mathrm{RN}}$ and $\gamma_{\mathrm{RN}}$), distances ($L_I$), and the individual optimal signal-to-noise ratios (SNRs, $\rho_I$) for the 20 pulsars in \textbf{Dataset A}.
Similarly, Table~\ref{tab:pulsar_snr_30} provides the corresponding parameters and SNRs for the 30 pulsars in \textbf{Dataset B}.

For both datasets, the white noise parameters were fixed for all pulsars: the root-mean-square (rms) timing residual was set to $100~\mathrm{ns}$ (corresponding to an EFAC parameter of $1.0$ and the $\log_{10}$ of the ECORR/EQUAD parameters set to $-8.0$). 
The pulsar distance priors were set to $L_I \in 1.2\times[0.8,\,1.2]~\mathrm{kpc}$ for all sources.
The individual SNRs are reported for the configuration with the strongest injected signal ($\rho_{\mathrm{tot}} \approx 30$), illustrating how differences in noise properties and geometric alignment lead to significant variations in sensitivity across the array.
\begin{table}[!htbp]
\centering
\setlength{\tabcolsep}{2.2pt}
\footnotesize
\caption{
\textbf{Dataset A (20 Pulsars):} Injected red noise parameters, distances, and individual SNRs.
The SNRs are calculated based on the dataset with the highest signal strength ($\rho_{\mathrm{tot}} \approx 30.40$).
}
\label{tab:pulsar_combined}
\renewcommand{\arraystretch}{1.05}
\begin{tabular}{lcccc}
\hline
\hline
Pulsar & $\log_{10} A_{\mathrm{RN}}$ & $\gamma_{\mathrm{RN}}$ & $L_I$ & $\rho_I$ \\
\hline
J1144$-$0287 & $-16.11$ & $3.07$ & \multirow{20}{*}{$1.2$} & $20.47$ \\
J1304$+$4859 & $-16.11$ & $2.52$ & & $11.41$ \\
J1354$-$3337 & $-16.81$ & $4.77$ & & $10.85$ \\
J0814$-$2049 & $-16.77$ & $2.74$ & & $6.93$ \\
J0604$+$0863 & $-15.05$ & $4.70$ & & $6.26$ \\
J2214$+$5821 & $-16.09$ & $2.69$ & & $6.13$ \\
J0114$-$7181 & $-16.35$ & $1.94$ & & $5.55$ \\
J1844$+$3337 & $-15.17$ & $3.71$ & & $5.53$ \\
J1724$-$1448 & $-14.83$ & $4.18$ & & $5.19$ \\
J0444$-$4054 & $-16.68$ & $1.67$ & & $4.77$ \\
J1514$+$1448 & $-15.66$ & $3.21$ & & $3.02$ \\
J2054$+$0287 & $-16.22$ & $4.36$ & & $2.78$ \\
J0724$+$7181 & $-14.77$ & $4.95$ & & $1.33$ \\
J0354$+$4054 & $-15.16$ & $2.99$ & & $1.25$ \\
J0234$-$0863 & $-16.66$ & $3.77$ & & $0.69$ \\
J1024$-$5821 & $-13.43$ & $4.03$ & & $0.52$ \\
J0934$+$2674 & $-13.13$ & $4.17$ & & $0.44$ \\
J1934$-$4859 & $-13.51$ & $4.96$ & & $0.18$ \\
J0024$+$2049 & $-13.75$ & $4.20$ & & $0.07$ \\
J2304$-$2674 & $-13.82$ & $3.76$ & & $0.07$ \\
\hline
\hline
\end{tabular}
\end{table}

\begin{table}[!htbp]
\centering
\setlength{\tabcolsep}{2.2pt}
\footnotesize
\caption{
\textbf{Dataset B (30 Pulsars):} Injected red noise parameters, distances, and individual SNRs.
The pulsars are sorted by SNR contribution; the total network SNR is approximately 30.
}
\label{tab:pulsar_snr_30}
\renewcommand{\arraystretch}{1.05}
\begin{tabular}{lcccc}
\hline
\hline
Pulsar & $\log_{10} A_{\mathrm{RN}}$ & $\gamma_{\mathrm{RN}}$ & $L_I$ & $\rho_I$ \\
\hline
J1354$-$0191 & $-15.67$ & $2.02$ & \multirow{30}{*}{$1.2$} & $15.54$ \\
J1024$-$1349 & $-15.76$ & $2.53$ & & $9.39$ \\
J0814$+$0574 & $-15.23$ & $2.88$ & & $9.13$ \\
J0904$-$5006 & $-16.95$ & $2.57$ & & $8.27$ \\
J1304$+$5644 & $-16.79$ & $3.67$ & & $7.67$ \\
J1234$-$3452 & $-16.84$ & $3.94$ & & $6.89$ \\
J0604$+$2568 & $-15.28$ & $4.09$ & & $6.58$ \\
J0934$+$3903 & $-15.54$ & $1.37$ & & $6.45$ \\
J1724$+$0959 & $-16.81$ & $4.98$ & & $5.79$ \\
J0724$+$7516 & $-14.87$ & $1.49$ & & $5.73$ \\
J2214$+$6416 & $-15.35$ & $1.63$ & & $5.39$ \\
J1814$-$4443 & $-15.18$ & $1.15$ & & $5.33$ \\
J1444$-$6416 & $-15.67$ & $1.41$ & & $4.93$ \\
J0654$-$2568 & $-14.59$ & $1.52$ & & $3.63$ \\
J1844$+$4443 & $-14.49$ & $2.86$ & & $3.17$ \\
J0354$+$5006 & $-15.46$ & $4.58$ & & $3.02$ \\
J0444$-$0574 & $-14.44$ & $3.11$ & & $1.54$ \\
J0024$+$3452 & $-14.47$ & $1.30$ & & $1.32$ \\
J0534$-$7516 & $-16.45$ & $3.76$ & & $1.26$ \\
J1934$-$0959 & $-14.61$ & $4.71$ & & $1.20$ \\
J2144$-$3000 & $-16.56$ & $1.72$ & & $1.00$ \\
J2054$+$2151 & $-16.18$ & $2.14$ & & $0.97$ \\
J0234$+$1349 & $-14.98$ & $4.90$ & & $0.83$ \\
J2354$-$5644 & $-16.22$ & $1.31$ & & $0.70$ \\
J1144$+$1746 & $-13.56$ & $1.06$ & & $0.55$ \\
J0324$-$3903 & $-14.20$ & $4.20$ & & $0.27$ \\
J1514$+$3000 & $-13.48$ & $2.91$ & & $0.23$ \\
J2304$+$0191 & $-14.25$ & $1.09$ & & $0.16$ \\
J1604$-$2151 & $-13.17$ & $3.61$ & & $0.10$ \\
J0114$-$1746 & $-13.41$ & $3.42$ & & $0.01$ \\
\hline
\hline
\end{tabular}
\end{table}
\end{document}